\documentclass[aps,prl,twocolumn,amsmath,amssymb,nofootinbib,superscriptaddress]{revtex4-2}
\usepackage{times}
\usepackage{braket}
\usepackage[pdftex]{graphicx}
\usepackage{dcolumn}
\usepackage{bm}
\usepackage{bbm}
\usepackage{amsmath}
\usepackage{indentfirst}
\usepackage[colorlinks,linkcolor=blue,anchorcolor=blue,citecolor=blue]{hyperref}
\usepackage[dvipsnames]{xcolor}
\usepackage{xcolor}

\colorlet{shadecolor}{gray!40}

\usepackage{algorithm}
\usepackage{algorithmicx}
\usepackage{algpseudocode}
\usepackage{verbatim}
\usepackage{makecell}
\usepackage[normalem]{ulem}
\usepackage{setspace}
\usepackage{float}
\usepackage{booktabs}
\usepackage{threeparttable}
\newtheorem{theorem}{Theorem}

\begin{document}
\title{Low-Overhead and High-Fidelity Preparation of Logical Non-Clifford States with Multi-Level Transversal Injection}

\author{Jiaxuan Zhang}
\email{zhangjx1@iai.ustc.edu.cn}
\affiliation{Institute of Artificial Intelligence, Hefei Comprehensive National Science Center, Hefei, Anhui, 230088, P. R. China}

\author{Tian-Hao Wei}
\affiliation{Laboratory of Quantum Information, University of Science and Technology of China, Hefei, 230026, China}

\author{Xi-Ning Zhuang}
\affiliation{Laboratory of Quantum Information, University of Science and Technology of China, Hefei, 230026, China}

\author{Zhao-Yun Chen}
\affiliation{Institute of Artificial Intelligence, Hefei Comprehensive National Science Center}

\author{Wei-Cheng Kong}
\affiliation{Origin Quantum Computing,  Hefei,  China}

\author{Yu-Chun Wu}
\email{wuyuchun@ustc.edu.cn}
\affiliation{Laboratory of Quantum Information, University of Science and Technology of China, Hefei, 230026, China}
\affiliation{Anhui Province Key Laboratory of Quantum Network, University of Science and Technology of China, Hefei, 230026, China}
\affiliation{Institute of Artificial Intelligence, Hefei Comprehensive National Science Center}

\author{Guo-Ping Guo}
\affiliation{Laboratory of Quantum Information, University of Science and Technology of China, Hefei, 230026, China}
\affiliation{Anhui Province Key Laboratory of Quantum Network, University of Science and Technology of China, Hefei, 230026, China}
\affiliation{Institute of Artificial Intelligence, Hefei Comprehensive National Science Center}
\affiliation{Origin Quantum Computing,  Hefei,  China}
\date{\today}

\begin{abstract}

Rotation gates are widely used in various quantum algorithms. To implement fault-tolerant rotation gates, state distillation or gate synthesis is typically employed. However, the overhead of these schemes scales rapidly with increasing Clifford hierarchy levels and required fidelity, limiting their use in large-scale quantum algorithms. In this letter, we propose a method termed multi-level transversal injection (MLTI) for preparing rotation states at arbitrary Clifford hierarchy levels, which achieves high fidelity while significantly reducing overhead compared to previous methods. Unlike the linear cost scaling of state distillation, the overhead of MLTI drops with Clifford hierarchy levels and then plateaus at higher levels, slashing the required space-time volume by up to several orders of magnitude. Additionally, we introduce a method for eliminating the off-diagonal terms of rotation states without extra overhead, unifying the metrics of infidelity and trace distance. These results render the overhead for high-precision rotation gates no longer prohibitive, and thus bring the broad adoption of large-scale quantum algorithms a critical step closer to reality.

\end{abstract}
\maketitle

\textit{Introduction}.---Quantum computing holds the potential to address problems intractable for classical computers~\cite{Feynman+1988+523+548,PhysRevLett.103.150502,365700,freedman2002simulation}. Rotation gates constitute a fundamental class of operations ubiquitous in various quantum algorithms, such as data-encoding~\cite{grover2002creatingsuperpositionscorrespondefficiently,PhysRevLett.100.160501,PhysRevLett.129.230504,10012045} and quantum simulation~\cite{poulin2014trotterstepsizerequired,https://doi.org/10.1002/qua.24856,Kivlichan2020improvedfault,Campbell_2022}. They are particularly crucial for solving real-world problems, owing to the ability to encode classical information into quantum states.

Performing large-scale quantum algorithms necessitates quantum error correction (QEC)~\cite{preskill1998reliable,gottesman1997stabilizer} and fault-tolerant quantum computing (FTQC)~\cite{548464,campbell2017roads}. Primary methods for achieving fault-tolerant rotation gates include gate synthesis~\cite{dawson2005solovaykitaevalgorithm,ross2016optimalancillafreecliffordtapproximation} and state distillation~\cite{meier2012magicstatedistillationfourqubitcode,PhysRevA.91.042315,Campbell_2016}. Gate synthesis approximates single-qubit rotation gates through sequences of $T$ and $H$ gates, while state distillation prepares high-fidelity rotation states, subsequently enabling rotations by gate teleportation. These approaches share the characteristic that costs escalate rapidly as the Clifford hierarchy~\cite{gottesman1999quantum} levels and fidelity requirements increase. This reality impedes the implementation of quantum algorithms demanding extensive high-fidelity rotation gates.

Recent studies propose a method for preparing arbitrary-angle rotation states on logical qubits, termed transversal injection ~\cite{gavriel2022transversalinjectionmethoddirect,choi2023faulttolerantnoncliffordstate,PhysRevX.15.021057}. It works by initializing rotation states on physical qubits, followed by stabilizer measurements and post-selection. Compared to gate synthesis or state distillation, transversal injection incurs significantly lower overhead. However, the fidelity of the output state in transversal injection is proportional to the physical error rate. High fidelity is achievable only when the rotation angle of the output state is extremely small. Consequently, current transversal injection methods are predominantly applicable to early-FTQC~\cite{PhysRevX.15.021057,PRXQuantum.5.020101}, yet impractical for large-scale algorithm implementations.

In this letter, we propose a method termed multi-level transversal injection (MLTI) to prepare high-fidelity rotation states. The key idea is to construct the transversal injection protocol on the logical qubit level through lattice surgery operations. We establish a theorem guaranteeing the continuous reduction of infidelity after each level of MLTI. Additionally, to address the rapid decrease in rotation angles after multiple levels, a technique called magic state pumping is introduced, enabling the output states to reside at arbitrary Clifford hierarchy level. We further propose a randomization method during gate teleportation, which eliminates diagonal terms of rotation states and thus unifies the infidelity and trace distance metrics. 

Numerical simulations demonstrate that MLTI holds a significant and increasing advantage over distillation or gate synthesis for implementing high-level rotation gates. For instance, when the Clifford hierarchy level \(l > 7\) at a physical error rate of \(5 \times 10^{-4}\), the space-time volume of MLTI for rotation state preparation with infidelity $<10^{-12}$ is consistently lower than that of state distillation. Particularly for cases with $l>12$, the volume is reduced by a factor of over 10. By using these rotation states to execute gate teleportation circuits, the overhead for implementing rotation gates across all Clifford hierarchy levels is significantly lowered, ranging from a several-fold to an orders-of-magnitude reduction, compared to gate synthesis. 

\textit{Review of physical-level transversal injection}.---First, we clarify some basic notations. All rotation gates in this work are restricted to the $Z$-basis, i.e., $R_z(\theta)=e^{i\theta Z}$. Correspondingly, a rotation state with angle $\theta$ is defined as
\begin{equation}
\begin{aligned}
\ket{\theta}=R_z(\theta)\ket{+}=\cos\theta \ket{+} + i \sin\theta \ket{-}.
\end{aligned}
\end{equation}
In particular, the gate with a rotation angle of $\pi/2^l$ belongs to the $l$-th level of the Clifford hierarchy~\cite{gottesman1999quantum}. For simplicity, we also refer to the rotation state $\ket{\pi/2^l}$ as belonging to the $l$-th Clifford hierarchy level.

The physical-level transversal injection~\cite{choi2023faulttolerantnoncliffordstate} starts from the logical state $\ket{+_L}$, followed by applying the rotation gates $R_z(\alpha)$ to all $k$ physical qubits supporting the logical operator $Z_L$. This state can be expanded as 
\begin{equation}
\begin{aligned}
&R_z(\alpha)^{\otimes k} \ket{+_L}
=(\cos\alpha I+ i\sin\alpha Z)^{\otimes k} \ket{+_L}\\
&=\sum_{P\in\{I,Z\}^{\otimes k}}(\cos\alpha)^{\text{wt}(P)} (i\sin\alpha)^{k-\text{wt}(P)} P \ket{+_L},
\end{aligned}\label{eq2}
\end{equation}
where $\text{wt}(P)$ is the weight of the Pauli operator $P$, i.e., the number of non-identity operators in the Pauli product. In the absence of noise, subsequent stabilizer measurements yield the syndrome of $P$, which is trivial only when $P = I^{\otimes k}$ or $Z_L$. Thus, after post-select the $+1$ syndrome, the ideal output state is 
\begin{equation}\label{eq3}
\begin{aligned}
\ket{\beta_L}=\frac{1}{p_s^{(0)}}(\cos ^{k}\alpha \ket{+_L} + i \sin^{k}\alpha \ket{{-}_L})
\end{aligned}
\end{equation}
where 
\begin{equation}\label{eq4}
\begin{aligned}
\beta =\arctan (\tan^{k}\alpha) \simeq \alpha^k
\end{aligned}
\end{equation}
is the angle of output rotation state and $p_s^{(0)}=\cos ^{2k}\alpha + \sin^{2k}\alpha$ is the rate of passing the post-selection. Here, we fix the phase of the second term to be $+i$, achievable by applying $Z_L$ or $S_L$. 
For simplicity, we denote the transverse injection process with input and output state rotation angles of $\alpha$ and $\beta$, and $k$ input rotation states, as $k\ket{\alpha}\xrightarrow{}\ket{\beta}$.

In the presence of noise, other terms in Eq.~(\ref{eq2}) may also pass post-selection, reducing the output fidelity. Assuming a physical error rate of $p_{phy}$, the infidelity of the output state scales as~\cite{PhysRevX.15.021057,SupplementaryInformation}
\begin{equation}
\begin{aligned}
\epsilon_L \propto  k \beta^{2(1-1/k)} p_{phy}+ \mathcal{O}(p_{phy}^2).  
\end{aligned}
\end{equation}

Following the initial version of transversal injection, a protocol based on multi-qubit rotation was proposed~\cite{PhysRevX.15.021057}. This approach significantly enhances the post-selection success rate, improving the practicality of transversal injection. To analyze this protocol, it suffices to let $k=d/m$ in the preceding analysis, where $d$ is the code distance and $m$ typically takes values of 2 or 3.  

Overall, the output fidelity of the transversal injection protocols exhibits two key characteristics. First, the infidelity of the output state decreases approximately quadratically as the angle $\beta$ diminishes. Second, the infidelity is fundamentally constrained by the physical error rate. High fidelity is achievable only when the output rotation angle $\beta$ is sufficiently small.

\textit{Logical-level transversal injection}.---To overcome the fidelity limitation imposed by the physical error rate, we consider executing transversal injection analogously on the logical qubit level. For convenience, we focus on transversal injection applied to surface code~\cite{PhysRevA.86.032324} logical qubits. Assume that $k$ logical states $\ket{\alpha}$ on surface codes have been prepared. Following the principle of transversal injection, terms containing $P={I}^{\otimes k}$ and ${P}={Z}^{\otimes k}$ in Eq.~(\ref{eq2}) must be identified.  

\begin{figure*}[t]
\centering
\includegraphics[width=17.5cm]{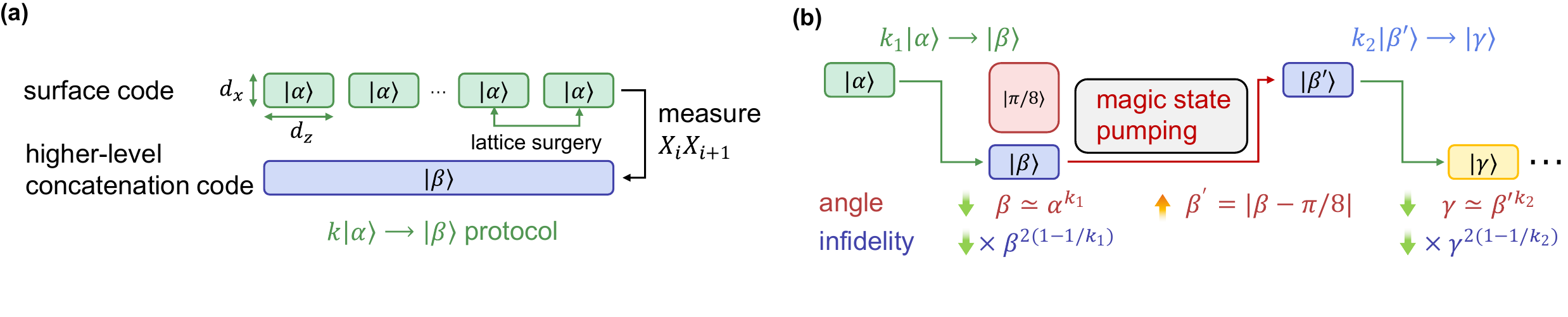}
\vspace{-0.6cm}
\caption{\textbf{Overview of the multi-level transversal injection protocol.}  
\textbf{(a)} Logical-level transversal injection. The transversal injection process on logical qubits is realized by constructing higher-level concatenated codes by lattice surgery operations to measure the operators \(X_i X_{i+1}\) $(i=1,2,...,k-1)$ on \(k\) surface code logical qubits.  
\textbf{(b)} The multi-level protocol continuously reduces the infidelity while maintaining the rotation angle of the output state from decreasing rapidly by magic state pumping.
}\label{fig1}
\end{figure*}

A key observation is that $P$ is detected by a higher-level code whose logical $Z$ operator is $Z^{\otimes k}$. For surface code logical qubits, error detection can be achieved by the higher-level concatenation code~\cite{nielsen2010quantum}. Specifically, to identify the syndrome of $P$, it suffices to measure the operators ${X}_i {X}_{i+1}$ ($i=1,2,...,k-1$), where ${X}_i$ denotes the logical $X$ operator on the $i$-th surface code logical qubit. This effectively results from concatenating surface codes with a repetition code (see Fig.~\ref{fig1}a).  

In practice, these operators can be measured by lattice surgery~\cite{Horsman_2012,Litinski2018latticesurgery,Litinski2019gameofsurfacecodes}. The desired concatenated code is constructed solely by executing the merging stage of lattice surgery. The subtlety is that these operators can be measured in parallel, as the readout regions for them do not overlap in lattice surgery~\cite{SupplementaryInformation}. 

After lattice surgery, only cases where all syndromes are $+1$ are preserved. Note that the concatenation code constitutes a larger-sized surface code. Depending on subsequent requirements, the size of this surface code can be expanded or reduced.

The ideal output state of logical-level transversal injection has the same form as Eq.~(\ref{eq3}). However, noise analysis becomes more complex because it involves decoding and error correction, fundamentally distinguishing it from the physical level. Critically, by assuming that all noise equivalently comes from the input states, we can establish a concise theorem that directly relates the input and output infidelity. 
\begin{theorem}
For a transversal injection protocol $k\ket{\alpha}\xrightarrow{}\ket{\beta}$ with small angle $\alpha$ applied on $k$ noisy input states $\rho$, the infidelity between the output state and target state $\ket{\beta_L}$ satisfies $\epsilon_L \leq k^2\beta^{2(1-1/k)} \epsilon+ \mathcal{O}(\epsilon^{1.5})$, where $\epsilon$ is the infidelity between $\rho$ and the ideal input state $\ket{\alpha}$. 
\end{theorem}
A detailed proof of Theorem~1 is provided in the Supplementary Material~\cite{SupplementaryInformation}. 

\textit{Multi-level transversal injection}.---Theorem~1 claims that the infidelity of the output state on each level of transversal injection is suppressed by a factor of $\beta^{2(1-1/k)}$. A naïve idea is iteratively executing multi levels of transversal injection to continuously reduce the output infidelity. However, the input-output relationship for the rotation angle in Eq.~(\ref{eq4}) indicates that the rotation angle decreases rapidly in multi-level protocols, limiting high-fidelity output states to extremely small angles.  

To counteract the persistent reduction in the rotation angle of the output state, we introduce a method termed magic state pumping (see Fig.~\ref{fig1}b). The core idea is to increasing the absolute value of the output state's rotation angle to near $\pi/8$ after each level of protocol. Since the rotation angle $\beta$ of the output state is typically much smaller than $\pi/8$, this can be achieved by applying $R_z(-\pi/8)$ and $X$ gates to the output state $\ket{\beta}$. 

\begin{figure}[tbp]
\centering
\includegraphics[width=8.5cm]{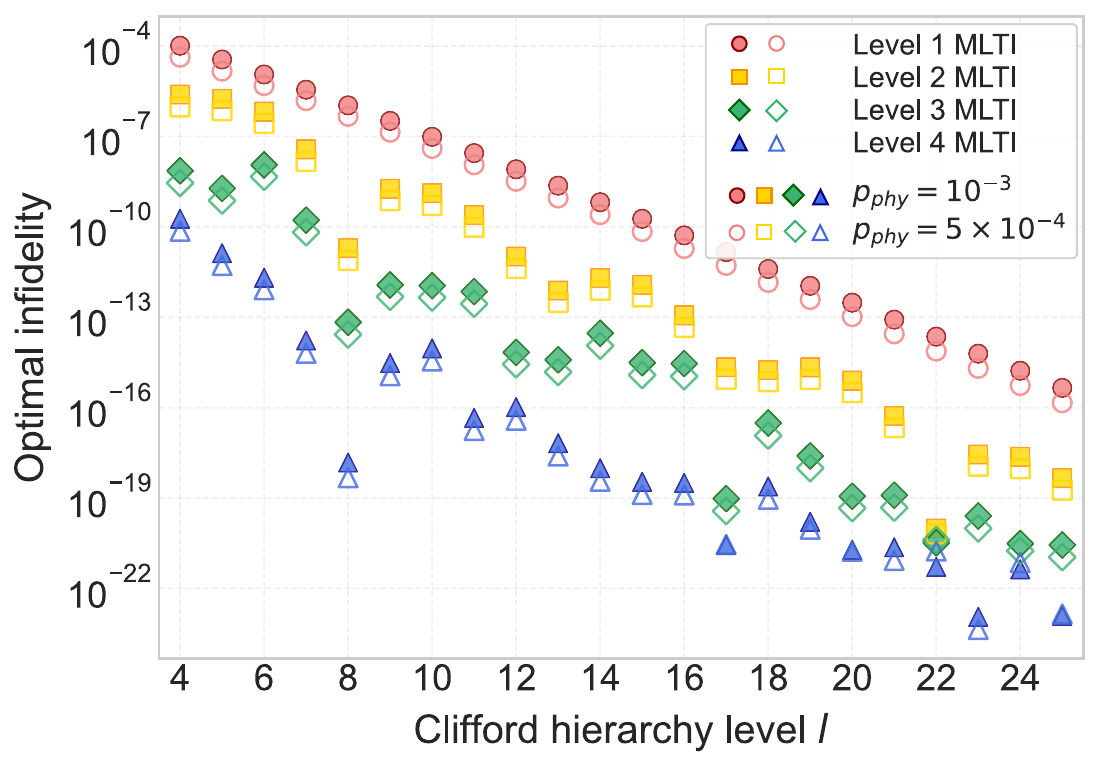}
\caption{\textbf{Optimal infidelity of the MLTI protocol on various levels at physical error rates of $10^{-3}$ or \(5 \times 10^{-4}\).} As the level of the MLTI protocol increases, the achievable lower bound of infidelity continues to decrease. 
}\label{fig2}
\end{figure}

Here, implementing a $R_z(-\pi/8)$ gate consumes a $\ket{\pi/8}$ state, which is precisely the well-known magic state~\cite{PhysRevA.71.022316} (up to a Pauli operator). To ensure that the noise in the $R_z(-\pi/8)$ gate is not significant, the fidelity of the magic state must be comparable to that of the output state. Fortunately, existing work~\cite{Litinski2019magicstate,Gidney2019efficientmagicstate,gidney2024magicstatecultivationgrowing} has extensively optimized methods for preparing high-fidelity magic states, making the overhead of magic state pumping not particularly large.

Now, the overall framework of the proposed MLTI protocol can be described. The first level of MLTI is the physical-level transversal injection. Subsequently, using magic states, ${\pi/8}$ rotation operations is performed on the output state to serve as the input on the next level. According to Theorem 1, the infidelity of the output state on each level is multiplied by a factor of $k^2 \beta^{2(1-1/k)}$ (at least), and thus continuously decreases as the protocol level increases.

Given the output state $\ket{\gamma}$ on the final level $r$ and the number of input states $k$, the input state $\ket{\beta'}$ on level $r$ can be derived from Eq.~(\ref{eq4}). Correspondingly, the rotation angle of the output state on level $r-1$ needs to be $\beta = \pi/8 - \beta'$. From this, the input state for the first level can be derived backwards.

Compared to physical-level transversal injection, MLTI significantly improves the fidelity upper bound of the protocol. Under physical-level transversal injection, for rotation states where the angle is not very small, the infidelity cannot be reduced sufficiently low. In contrast, under MLTI, the optimal infidelity continuously decreases as the level increases (see Fig.~\ref{fig2}). Here, the optimal infidelity is based on the premise that the only noise source is the physical error rate $p_{phy}$ of the first-level protocol, because subsequent noise can be suppressed through sufficiently large code distances and magic state distillation.

Furthermore, considering minimizing the overhead for a target infidelity, we use space-time volume as the metric. The space-time volume for each level of the protocol is defined as the product of physical qubit count and QEC cycles required, on average, to produce one target state. This volume comprises three components: executing the transversal injection protocol, preparing magic states, and performing $R_z(-\pi/8)$ gates. The total overhead of MLTI is the sum across all levels.

To reduce overhead, two minor optimizations are made to the implementation of MLTI. First, we employ rectangular rotated surface codes with asymmetric distances ($d_z > d_x$), since $Z$ errors more severely affect infidelity. Specifically, a $Z$ error flips $\ket{\theta}$ to the orthogonal state, while an $X$ error only contributes $\sin^2 2\theta$ to the infidelity. Second, we adapt code distances on each level of the protocol by patch deformation\cite{Litinski2019gameofsurfacecodes}, scaling the output state to a suitable size.

By searching for the number of input states $k$, the code distances $d_z$ and $d_x$, and the required infidelity of the magic state on each level, the space-time volume for the target infidelity can be minimized. The Supplementary Material~\cite{SupplementaryInformation} provides numerical methods for calculating infidelity and details on search pruning. To compare with the distillation protocol, we designed a surface-code implementation of the distillation protocol~\cite{SupplementaryInformation} to evaluate its overhead.

\begin{figure}[tbp]
\centering
\includegraphics[width=8.5cm]{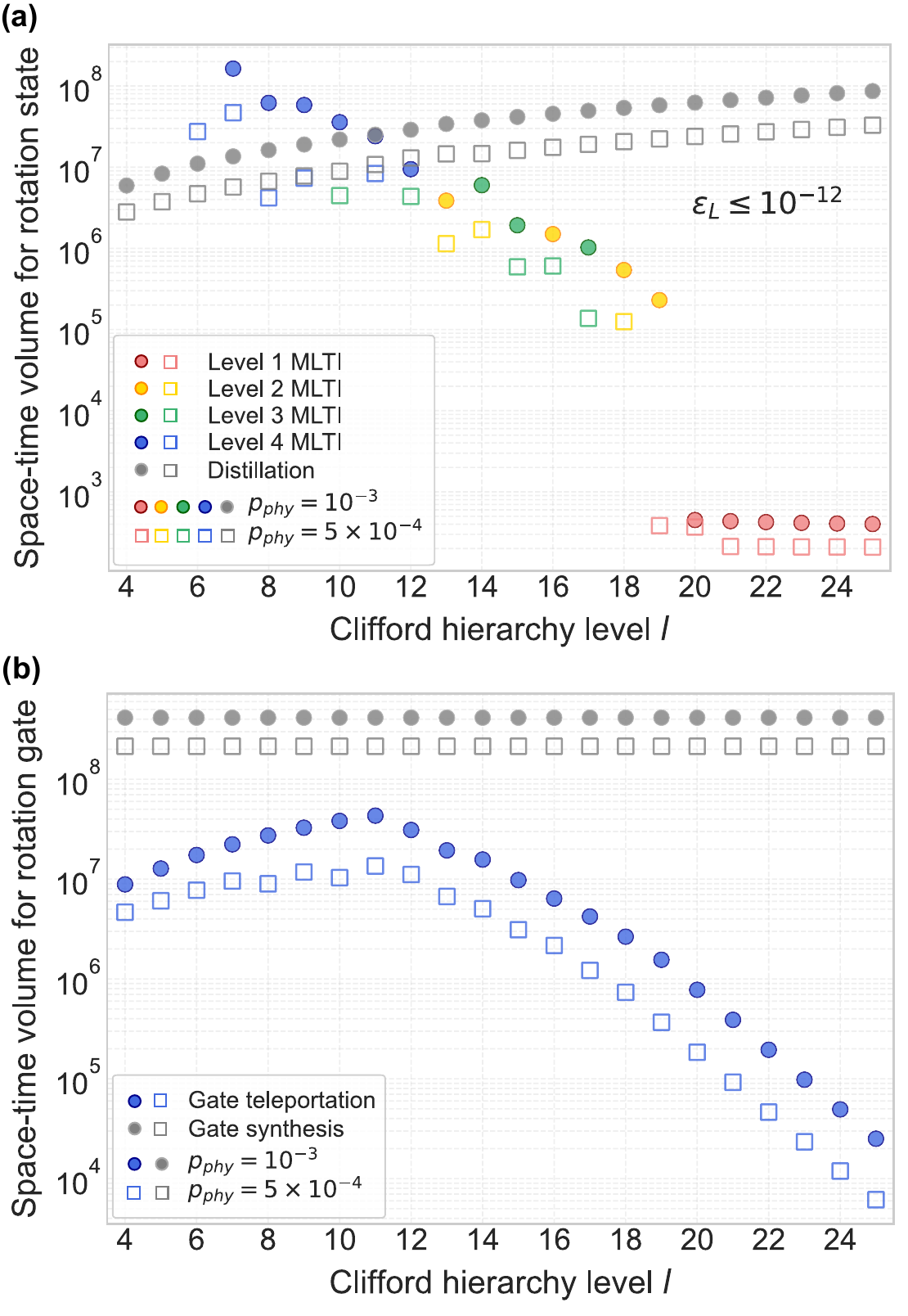}
\caption{\textbf{Comparison between MLTI and state distillation or gate synthesis protocols at physical error rates of \(10^{-3}\) or \(5 \times 10^{-4}\). } 
\textbf{(a)} Space-time volume required to prepare rotation states at Clifford hierarchy levels from 4 to 25 using MLTI and state distillation, with a target infidelity of \(10^{-12}\). For the MLTI results, only the least space-time volume among protocols on level 1-4 is presented.
\textbf{(b)} Space-time volume required to implement rotation gates $R_z(\pi/2^l)$ by gate teleportation using the rotation states with least overhead from (a), and the space-time volume required by gate synthesis to achieve same rotation gates with comparable error rates.
}\label{fig3}
\end{figure}

Fig.~\ref{fig3}a presents numerical results of the space-time volumes for preparing rotation states at different Clifford hierarchy levels $l$ using MLTI and state distillation~\cite{Campbell_2016}, with a target infidelity of $10^{-12}$ and physical error rates of $5\times 10^{-4}$ or $10^{-3}$. Fig.~\ref{fig3}b shows the space-time volumes required for implementing rotation gates $R_z(\pi/2^l)$ by gate teleportation using these rotation states, along with the overhead of magic state consumption by gate synthesis~\cite{ross2016optimalancillafreecliffordtapproximation} under comparable gate error rates for comparison. Unsurprisingly, the overhead of state distillation increases approximately linearly with $l$, and the gate synthesis overhead remains consistently high, due to the fixed target infidelity. In contrast, MLTI follows a completely opposite trend, whose overhead decreases with increasing $l$ and eventually plateaus at a negligible level. This reduction leads to a decrease in overhead by factors ranging from several times to several orders of magnitude for gate implementation. Furthermore, compared to the original transversal injection protocol~\cite{choi2023faulttolerantnoncliffordstate,PhysRevX.15.021057}, MLTI with up to level $r=4$ enables high-fidelity preparation of all rotation states with $l>6$, which is not achievable with the physical-level transversal injection protocol.

\textit{Elimination of off-diagonal terms in rotation states}.---We have demonstrated that  MLTI can generate rotation states with high fidelity. However, fidelity as a metric cannot assess the effect of off-diagonal errors. A more stringent metric is trace distance~\cite{nielsen2010quantum}, which is directly related to the diamond distance~\cite{nielsen2010quantum} of the rotation gate, satisfying the chaining property and stability~\cite{PhysRevA.71.062310}. If a single-qubit state contains only diagonal terms in the basis of the target state and its orthogonal state, infidelity equals trace distance.

For a target state $\ket{\theta}$, one method to eliminate off-diagonal terms is to apply the $R_z(2\theta)X$ gate with probability $1/2$, corresponding to the following channel
\begin{equation}
\begin{aligned}
\mathcal{D}(\rho) = \frac{1}{2}(\rho + R_z(2\theta)X \rho X R_z^\dagger(2\theta)).
\end{aligned}
\end{equation}
However, this introduces the additional cost of executing $R_z(2\theta)$.

Our key idea is to directly incorporate this randomness into the gate teleportation circuit, avoiding the extra execution of $R_z(2\theta)$. Specifically, an $X$ gate is applied to the rotation state $\ket{\theta}$ with probability $1/2$, followed by gate teleportation circuit. If the $X$ gate was applied, the measurement feedback condition is modified to ensure that the gate-teleportation still performs $R_z(\theta)$.

The following brief discussion illustrates how this method eliminates off-diagonal terms. Assume the gate-teleportation channel is $\mathcal{G}$, satisfying $\mathcal{G}(\rho_{in} \otimes \ket{\theta}\bra{\theta}) = \rho_{out} \otimes I$, where $\rho_{out} = R_z(\theta)\rho_{in}R_z(\theta)^\dagger$. Similarly, by modifying the measurement feedback condition, there exists a channel $\mathcal{G}'$ satisfying $\mathcal{G}'(\rho_{in} \otimes X\ket{\theta}\bra{\theta}X) = \rho_{out} \otimes I$. Now consider the effect of $\mathcal{G}$ and $\mathcal{G}'$ on the off-diagonal term $\ket{\theta^\perp}\bra{\theta} = Z\ket{\theta}\bra{\theta}$. Commuting the $Z$ operator out of $\mathcal{G}$ or $\mathcal{G}'$, we have 
\begin{equation}
\begin{aligned}
&\mathcal{G}(\rho_{in} \otimes \ket{\theta^\perp}\bra{\theta}) = Z \otimes Z \rho_{target} \otimes I,
\\
&\mathcal{G}'(\rho_{in} \otimes X\ket{\theta^\perp}\bra{\theta}X) = -Z \otimes Z \rho_{target} \otimes I.
\end{aligned}
\end{equation} 
Therefore, by executing $\mathcal{G}$ or $\mathcal{G}'$ each with probability $1/2$, the effects of the off-diagonal term cancel out. The complete proof of this is provided in the Supplementary Material~\cite{SupplementaryInformation}. Similarly, this randomness can be introduced in subsequent correction operations $R_z(2\theta)$, $R_z(4\theta)$, etc., thereby completely avoiding off-diagonal terms in the executions of rotation gates.

\textit{Discussion}.---Rotation gates are ubiquitous in quantum algorithms. In applications such as state preparation~\cite{grover2002creatingsuperpositionscorrespondefficiently,PhysRevLett.100.160501}  or block encoding~\cite{PhysRevLett.129.230504,10012045}, the overhead for fault-tolerant implementation of rotation states generally dominates~\cite{zhuang2025pathwaypracticalquantumadvantage}. The reduced overhead underlines the potential of MLTI as a key enabler for implementing quantum algorithms that require numerous high-precision rotations, a task that would be prohibitively expensive using state distillation or gate synthesis. 

We also remarked that MLTI is not restricted to a specific QEC code. For platforms with more flexible connectivity, such as neutral atoms~\cite{bluvstein2024logical} or ion traps~\cite{physrevx.11.041058}, considering high-rate quantum low-density-parity check (QLDPC) codes~\cite{PRXQuantum.2.040101,6671468,bravyi2024high} in the underlying encoding and higher-level concatenation codes could further reduce overhead. We leave this exploration to future work.

\begin{acknowledgments}
This work was supported by the National Key Research and Development Program of China (Grant No. 2024YFB4504101 and Grant No. 2023YFB4502500) and Anhui Province Innovation Plan for Science and Technology (Grant No. 202423r06050002). 
\end{acknowledgments}

\bibliographystyle{apsrev4-2}
\bibliography{ref}

\clearpage
\setcounter{table}{0}
\renewcommand{\thetable}{S\arabic{table}}%
\setcounter{figure}{0}
\renewcommand{\thefigure}{S\arabic{figure}}%
\setcounter{section}{0}
\setcounter{equation}{0}
\renewcommand{\theequation}{S\arabic{equation}}%

\onecolumngrid
\begin{center}
{\large \bf Supplemental Material for ``Low-Overhead and High-Fidelity Preparation of Logical Non-Clifford States with Multi-Level Transversal Injection''}
\vspace{0.3cm}
\end{center}

\setcounter{page}{1}

\section{Review of Several Protocols for Implementing Rotation Gates}
\subsection{Physical-Level Transversal Injection}

The physical-level transversal injection~\cite{choi2023faulttolerantnoncliffordstate} starts from the $\ket{+}$ state on all $n$ data qubits of the QEC code. Here, we only focus on CSS codes. Then, rotation gates $R_z(\alpha)$ are applied to $k$ physical qubits constituting $Z_L=Z_1Z_2..Z_{k}$. Subsequently, stabilizer measurements are performed. This process differs slightly from our description in the main text. The following equation demonstrates that it is equivalent to first applying $R_Z^{\otimes k}(\alpha)$ gates to the $\ket{+_L}$ state, followed by stabilizer measurements:
\begin{equation}
\begin{aligned}
\Pi_s R_z^{\otimes k}(\alpha)\ket{+}^{\otimes n} 
=\Pi_x \Pi_z R_z^{\otimes k}(\alpha)\ket{+}^{\otimes n}
=\Pi_x R_z^{\otimes k}(\alpha) \Pi_z \ket{+}^{\otimes n}
=\Pi_x R_z^{\otimes k}(\alpha) \ket{+_L},
\end{aligned}
\end{equation}
where $\Pi_s = \Pi_x \Pi_z$, and $\Pi_x$ and $\Pi_z$ are the $X$-type and $Z$-type measurement operators corresponding to the stabilizer generators of the QEC code, respectively. 

The state $R^{\otimes k}(\alpha)\ket{+_L}$ can be expanded as
\begin{equation}
\begin{aligned}\label{s1}
R^{\otimes k}(\alpha)\ket{+_L}
=(\cos\alpha I+ i\sin\alpha Z)^{\otimes k} \ket{+_L}
=\sum_{P\in\{I,Z\}^{\otimes k}}(\cos\alpha)^{\text{wt}(P)} (i\sin\alpha)^{k-\text{wt}(P)} P \ket{+_L},
\end{aligned}
\end{equation}
where $\text{wt}(P)$ is the weight of the Pauli operator $P$, i.e., the number of non-identity operators in the Pauli product. In the absence of noise, the subsequent $\Pi_x$ measurements will yield the error syndrome of $P$. The syndromes are trivial only when $P=I^{\otimes k}$ or $P=Z_L=Z^{\otimes k}$. Therefore, after post-selecting on the $+1$ error syndromes, the resulting quantum state is:
\begin{equation}
\begin{aligned}
\frac{1}{p_s^{(0)}}(\cos ^{k}\alpha \ket{+_L} + i^{k}\sin^{k}\alpha \ket{{-_L}}),
\end{aligned}\label{output_state}
\end{equation}
where the rate of passing the post-selection is
\begin{equation}
\begin{aligned}
p_s^{(0)}=\cos ^{2k}\alpha + \sin^{2k}\alpha \simeq 1-k\alpha^{2}+\mathcal{O}(\alpha^4).
\end{aligned}
\end{equation}
Finally, the phase $i^{1-k}$ of the second term is compensated. This can be achieved by applying $Z_L$ or $S_L$ to Eq.~(\ref{output_state}). The ideal output state $\ket{\psi_0}$ of the protocol is defined as
\begin{equation}
\begin{aligned}
\ket{\psi_0}\equiv \cos \beta \ket{+_L} + i\sin\beta \ket{{-_L}}
=\frac{1}{p_s^{(0)}}(\cos ^{k}\alpha \ket{+_L} + i \sin^{k}\alpha \ket{{-}_L}) =\ket{\beta_L} 
\end{aligned}
\end{equation}
where
\begin{equation}
\begin{aligned}
\beta =\arctan (\tan^{k}\alpha ) \simeq \alpha^{k}
\end{aligned}
\end{equation}

In the noisy case, the operator $P$ in Eq.~(\ref{s1}) might not be detectable. Define the following state
\begin{equation}\label{s7}
\begin{aligned}
\ket{\psi_m}
=\frac{1}{p_s^{(m)}}(\cos ^{k-m}\alpha \sin ^{m}\alpha \ket{+_L} + i^{1-m}\sin^{k-m}\alpha \cos ^{m}\alpha \ket{{-}_L}),
\end{aligned}
\end{equation}
where
\begin{equation}
\begin{aligned}
p_s^{(m)}=\cos ^{2k-2m}\alpha \sin^{2m}\alpha + \sin^{2k-2m}\alpha \cos ^{2m}\alpha.
\end{aligned}
\end{equation}
This state corresponds to the terms in Eq.~(\ref{s1}) where $\text{wt}(P)=m$ and $\text{wt}(P)=k-m$. The state obtained through post-selection is $\ket{\psi_m}$ when the terms containing operators $P$ and $PZ_L$ in Eq.~(\ref{s1}) yield a trivial error syndrome.
This occurs under following two scenarios. First, an error $P$ or $PZ_L$ occurs exactly on the data qubits, which precisely cancels the error syndromes of $P$ or $PZ_L$. Alternatively, errors occur on the syndrome qubits, causing the syndrome to be misread. On the surface code, using the method from Ref.~\cite{PhysRevX.15.021057}, syndrome extraction is performed twice, and post-selection is applied only to the $X$-type stabilizers in the first two rows of the surface code array, retaining outcomes where both rounds yield +1. Under this setup, the probability of measurement errors can be suppressed to $\mathcal{O}(p_{phy}^2)$. Therefore, it is sufficient to consider the probability of errors $P$ or $PZ_L$ occurring on the data qubits if we only care about terms up to $\mathcal{O}(p_{phy})$. In fact, let $p_e^{(m)}$ denote the probability of $Z$ errors occurring equivalently on $m$ out of the $k$ physical qubits, then $p_e^{(m)}=\mathcal{O}(p_{phy}^m)$.

Overall, the output quantum state of the MI protocol in the noisy case can be written in mixed-state form as:
\begin{equation}
\begin{aligned}
\rho_{out}&=\frac{1}{p_s}((1-\sum_{m\geq1} p_e^{(m)})p_s^{(0)}\ket{\psi_0}\bra{\psi_0}+\sum_{m\geq1}p_e^{(m)}p_s^{(m)}\ket{\psi_m}\bra{\psi_m})\\
&\simeq \frac{1}{p_s}((1-p_e^{(1)})p_s^{(0)}\ket{\psi_0}\bra{\psi_0}+p_e^{(1)}p_s^{(1)}\ket{\psi_1}\bra{\psi_1}) + \mathcal{O}(p_{phy}^2)
\end{aligned}
\end{equation}
where $p_s$ is the probability of passing post-selection in the noisy case, satisfying:
\begin{equation}
\begin{aligned}
p_s=(1-\sum_{m\geq1} p_e^{(m)})p_s^{(0)}+\sum_{m\geq1}p_e^{(m)}p_s^{(m)}
\end{aligned}
\end{equation}
To calculate the infidelity $\epsilon_L$ between $\rho_{out}$ and the ideal output state $\ket{\psi_0}$, it is useful to consider the state orthogonal to $\ket{\psi_0}$:
\begin{equation}
\begin{aligned}
\ket{\psi_0^\perp} = \frac{1}{p_s^{(0)}}(i \sin^{k}\alpha \ket{+_L} + \cos ^{k}\alpha \ket{{-}_L}).
\end{aligned}
\end{equation} 
Then the infidelity can be computed by converting it to the fidelity between $\rho_{out}$ and $\ket{\psi_0^\perp}$, as follows:
\begin{equation}
\begin{aligned}
\epsilon_L 
&= \bra{\psi_0^\perp} \rho \ket{\psi_0^\perp}
\simeq \frac{p_e^{(1)}p_s^{(1)}}{p_s}|\braket{\psi_0^\perp|\psi_1}|^2 + \mathcal{O}(p_{phy}^2)\\
&=\frac{p_e^{(1)}}{p_sp_s^{(0)}}(\cos^{2(k-1)}\alpha\sin^{2(k+1)}\alpha+\sin^{2(k-1)}\alpha\cos^{2(k+1)}\alpha) + \mathcal{O}(p_{phy}^2)\\
&\simeq f(k)p_{phy}\alpha^{2(k-1)} + \mathcal{O}(p_{phy}^2) \\
&\simeq f(k)p_{phy}\beta^{2(1-1/k)}+ \mathcal{O}(p_{phy}^2) ,
\end{aligned}
\end{equation} 
where $f(k)$ is a linear function of $k$ and $\alpha$ is assumed to be a small angle.

For the multi-rotation MI protocol~\cite{PhysRevX.15.021057}, the data qubits in the logical operator $Z_L$ are partitioned into $k$ groups, and a multi-qubit rotation gate is performed on each group of data qubits. The only difference in the above discussion for the multi-rotation MI protocol is the need to modify the definition of $p_e^{(m)}$. In this case, single-qubit (or multi-qubit) errors might be detected, causing the post-selection to fail. Therefore, $p_e^{(m)}$ is defined as the probability of $Z$ errors occurring equivalently on $m$ out of the $k$ groups of physical qubits.

\subsection{State Distillation}
The state distillation protocol discussed here is based on the 4-qubit code, originally from Ref.~\cite{meier2012magicstatedistillationfourqubitcode}. Subsequently, Ref.~\cite{PhysRevA.91.042315} and \cite{Campbell_2016} proposed a more specific protocol implementation and optimization process for rotation states. Following these literatures, we consider here the rotation operation $R_y(\theta_l)$ around the $Y$-axis and the corresponding rotation state $\ket{Y_l}$, satisfying
\begin{equation}
\begin{aligned}
R_y(\theta_l)=e^{i\theta_l Y} = e^{i\pi Y/2^l},\quad \ket{Y_l} = R_y(\theta_l) \ket{+}=
\cos \frac{\pi}{2^l} \ket{+} + \sin \frac{\pi}{2^l} \ket{-},
\end{aligned}
\end{equation}
where $l$ denotes the Clifford hierarchy level. Particularly, note that $\ket{Y_2} = \ket{0}$. These states can be converted to the rotation gates or rotation states discussed in the main text by the Clifford operation $HS$.

\begin{figure}[t]
\centering
\includegraphics[width=18cm]{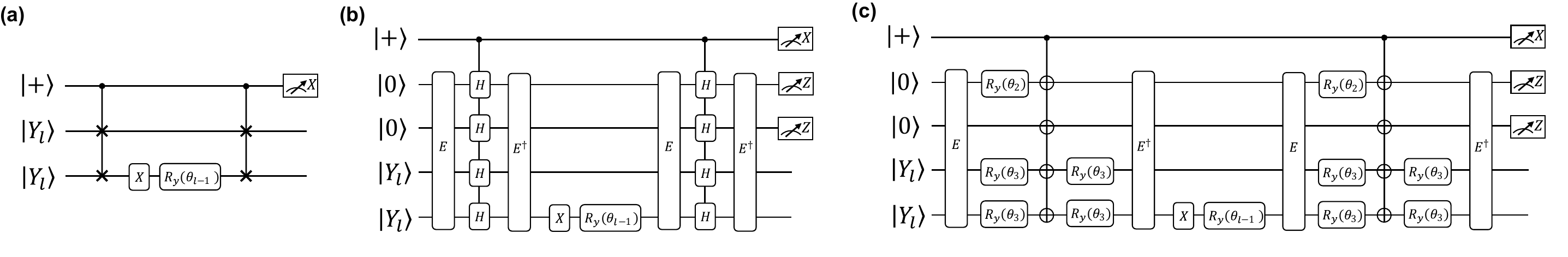}
\caption{\textbf{Distillation protocol applicable to rotation states.  }
\textbf{(a)} Primal circuit of the distillation protocol. Two noisy input states \(|\bar{Y}_l\rangle\) with error rate $\mathcal{O}(\epsilon)$ yield two distilled states with error rate $\mathcal{O}(\epsilon^2)$, requiring the execution of rotation gates at Clifford level \(l-1\).  
\textbf{(b)} Distillation circuit based on the \([[4,2,2]]\) code. The key is to implement a CSWAP gate within the logical space of the \([[4,2,2]]\) code by controlled-$H$ gates. \(E\) denotes the encoding circuit of the \([[4,2,2]]\) code.  
\textbf{(c)} Further simplification of the distillation circuit, adapted from Ref.~\cite{Campbell_2016}.
}\label{figs1}
\end{figure}

The basic circuit form of this distillation protocol is given in Fig~\ref{figs1}a. Note that
\begin{equation}
\begin{aligned}
 R_y(\theta_{l-1}) X = \ket{Y_l}\bra{Y_l}-\ket{Y_l^\perp}\bra{Y_l^\perp},
\end{aligned}
\end{equation}
where $Y_l^\perp =  R_y(\theta_l) \ket{-}$ is the state orthogonal to $\ket{Y_l}$. The input states on the second and third qubits are
\begin{equation}
\begin{aligned}
\tilde{\rho}^{\otimes2} = [(1-\epsilon)\ket{Y_l}\bra{Y_l} + \epsilon\ket{Y_l^\perp}\bra{Y_l^\perp}]^{\otimes2} .
\end{aligned}
\end{equation}
The post-selection condition is that the measurement outcome of the first qubit is $+1$. It is easy to verify that only the terms $\ket{Y_l}\bra{Y_l} \otimes \ket{Y_l}\bra{Y_l}$ and $\ket{Y_l^\perp}\bra{Y_l^\perp} \otimes \ket{Y_l^\perp}\bra{Y_l^\perp}$ can pass the post-selection. Therefore, the output state is
\begin{equation}
\begin{aligned}
\rho_{out} \propto (1-\epsilon)^2\ket{Y_l}\bra{Y_l} \otimes \ket{Y_l}\bra{Y_l} +
\epsilon^2 \ket{Y_l^\perp}\bra{Y_l^\perp} \otimes \ket{Y_l^\perp}\bra{Y_l^\perp}.
\end{aligned}
\end{equation}
The infidelity is suppressed from $\mathcal{O}(\epsilon)$ of the input state to $\mathcal{O}(\epsilon^2)$.

To fault-tolerantly implement the CSWAP gate in the circuit, we consider the [[4,2,2]] quantum code, whose logical space is defined by the stabilizer group
\begin{equation}
\begin{aligned}
<X_1X_2X_3X_4,\quad Z_1Z_2Z_3Z_4>, 
\end{aligned}
\end{equation}
and the logical operators are
\begin{equation}
\begin{aligned}
\bar{Z}_1=Z_1 Z_3, \quad \bar{X}_1=X_1 X_4,\\
\bar{Z}_2=X_1 X_3, \quad \bar{X}_2=Z_1 Z_4.
\end{aligned}
\end{equation}
Note that the two sets of logical operators are completely symmetric, so the SWAP gate within the logical space is simply $H_1H_2H_3H_4$. Based on this, the CSWAP gate can be implemented using controlled-$H$ gates. Furthermore, we have $H=R_y(\theta_3) X R_y(\theta_3)^\dagger$. Utilizing these relations, Ref.~\cite{PhysRevA.91.042315} obtained the distillation circuit shown in Fig~\ref{figs1}b. Subsequently, Ref.~\cite{Campbell_2016} further simplified this circuit to arrive at the final distillation circuit shown in Fig~\ref{figs1}c.

\subsection{Gate Synthesis}
To compare with the results in the main text, we also consider the gate synthesis approach. The core idea of gate synthesis is to approximate continuous-angle rotation gates using discrete basic gates (including $H, S, T$ gates). 

The Solovay-Kitaev theorem~\cite{dawson2005solovaykitaevalgorithm} provides the theoretical foundation for gate synthesis. It states that for any single-qubit unitary operation $U$ and any precision $\delta > 0$, there exists a sequence composed of Clifford+$T$ gates such that the synthesized gate $U'$ satisfies $||U - U'|| \leq \delta$. More importantly, the length of this sequence grows polynomially with $\log(1/\delta)$.

In practice, the Ross-Selinger algorithm~\cite{ross2016optimalancillafreecliffordtapproximation}, which is more efficient than the original Solovay-Kitaev method, is typically used. It provides an optimal Clifford+$T$ sequence for approximating the $R_z(\theta)$ gate, with a $T$-count of $n_T = 3\log_2(1/\delta)+\mathcal{O}(\log(\log(1/\delta)))$.

\section{Proof of Theorem 1 and Off-Diagonal Terms Eliminating Method}
\subsection{Proof of Theorem 1}
Consider the transversal injection protocol $k\ket{\alpha}\xrightarrow{}\ket{\beta_L}$. In the basis $\{\ket{\alpha}, \ket{\alpha^\perp}\}$, assume the noisy input state $\rho$ on each qubits is given by 
\begin{equation}
\begin{aligned}
\rho= (1-\epsilon) \ket{\alpha}\bra{\alpha} + \epsilon \ket{\alpha^\perp}\bra{\alpha^\perp} + b\ket{\alpha}\bra{\alpha^\perp} + b^*\ket{\alpha^\perp}\bra{\alpha},
\end{aligned}
\end{equation} 
where $1-\epsilon = \bra{\alpha}\rho\ket{\alpha}$ is the fidelity between $\rho$ and the ideal input state $\ket{\alpha}$. Since $\rho$ is positive semi-definite, we have $|b|^2\leq \epsilon -\epsilon ^2$.

Assume $\ket{\psi_0}$, $\ket{\psi_1}$ has the same form as Eq.~(\ref{s7}). Now, consider the leading terms of $\epsilon$ in $\rho^k$ after the stabilizer measurements in the MI protocol.   The zeroth-order term in $\epsilon$ is 
\begin{equation}
\begin{aligned}
\Pi_s (1-\epsilon)^k \ket{\alpha^{\otimes k}}\bra{\alpha ^{\otimes k}} \Pi_s = (1-\epsilon)^k p_s^{(0)}\ket{\psi_0}\bra{\psi_0},
\end{aligned}
\end{equation} 
which contributes 0 to the infidelity. Secondly, there are $k$ first-order terms of
\begin{equation}
\begin{aligned}
\Pi_s \epsilon(1-\epsilon)^{k-1} \ket{\alpha^{\otimes k-1}\alpha^\perp}\bra{\alpha ^{\otimes k-1}\alpha^\perp} \Pi_s = (1-\epsilon)^{(k-1)}\epsilon p_s^{(1)}\ket{\psi_1}\bra{\psi_1},
\end{aligned}
\end{equation}
with total contribution to the infidelity of  
\begin{equation}
\begin{aligned} 
k\epsilon(1-\epsilon)^{k-1}p_s^{(1)}|\braket{\psi_1|\psi_0^\perp}|^2.
\end{aligned}
\end{equation}

It is also necessary to consider the leading terms involving $b$ or $b^*$. For example, there are $k$ first-order terms 
\begin{equation}
\begin{aligned}
&\Pi_s (1-\epsilon)^{k-1} (b\ket{\alpha^{\otimes k}}\bra{\alpha ^{\otimes k-1}\alpha^\perp} 
+b^*\ket{\alpha^{\otimes k-1}\alpha^\perp}\bra{\alpha ^{\otimes k-1}})\Pi_s \\
= &(1-\epsilon)^{(k-1)}\sqrt{p_s^{(0)}p_s^{(1)}} (b\ket{\psi_0}\bra{\psi_1}+b^*\ket{\psi_1}\bra{\psi_0}).
\end{aligned}
\end{equation} 
However, these terms contribute 0 to the infidelity, since $\bra{\psi_0^\perp}\ket{\psi_0}=0$. Furthermore, there are $2\times\binom{k}{2}$ second-order terms \begin{equation}
\begin{aligned}
&\Pi_s (1-\epsilon)^{k-2} |b|^2\ket{\alpha^{\otimes k-2} \alpha\alpha^\perp}\bra{\alpha^{\otimes k-2} \alpha^\perp\alpha} 
\Pi_s \\
= &(1-\epsilon)^{k-2} |b|^2{p_s^{(1)}} \ket{\psi_1}\bra{\psi_1},
\end{aligned}
\end{equation} 
with contribution to the infidelity of  
\begin{equation}
\begin{aligned}
2\times\binom{k}{2}\times (1-\epsilon)^{k-2} |b|^2 p_s^{(1)}|\braket{\psi_1|\psi_0^\perp}|^2.
\end{aligned}
\end{equation} 

Since $|b|^2\leq \epsilon -\epsilon ^2$, the remaining terms are at least $\mathcal{O}(\epsilon^{1.5})$ or higher order. Combining the above results and accounting for the output state's normalization coefficient $\simeq 1$ with a small angle $\alpha$, the infidelity $\epsilon_L$ of the output state satisfies 
\begin{equation}
\begin{aligned}
\epsilon_L 
&= k\epsilon(1-\epsilon)^{k-1}p_s^{(1)}|\bra{\psi_1}\ket{\psi_0^\perp}|^2 + 2\times\binom{k}{2}\times (1-\epsilon)^{k-2} |b|^2 p_s^{(1)}|\bra{\psi_1}\ket{\psi_0^\perp}|^2 +\mathcal{O}(\epsilon^{1.5})\\
&\leq k^2 p_s^{(1)}|\bra{\psi_1}\ket{\psi_0^\perp}|^2 \epsilon+ \mathcal{O}(\epsilon^{1.5}) \\&\simeq k^2\beta^{2(1-1/k)} \epsilon + \mathcal{O}(\epsilon^{1.5}).
\end{aligned}
\end{equation}

\subsection{Proof of Off-Diagonal Terms Eliminating Method}
\begin{figure}[tpb]
\centering
\includegraphics[width=8cm]{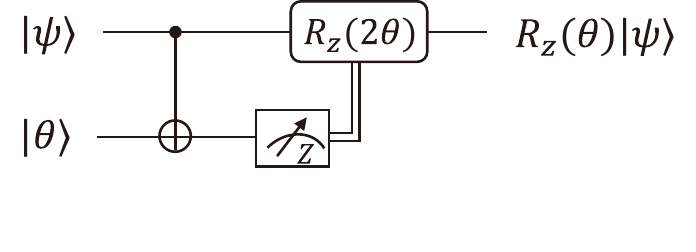}
\vspace{-0.7cm}
\caption{\textbf{Gate teleportation circuit for implementing the \(R_z(\theta)\) gate.} Based on the measurement outcome of the rotation state, a feedback operation is applied to decide whether to perform \(R_z(2\theta)\).
}\label{figs2}
\end{figure}

In this section, we provide a detailed proof that the randomization method in the main text eliminates the effect of off-diagonal terms in rotation states. Consider the gate teleportation circuit in Fig.~\ref{figs2}. The fundamental function of the gate teleportation circuit is that, when the input state is $\rho_{in} \otimes \ket{\theta}\bra{\theta}$, the output state is $\rho_{out} \otimes I$, where $\rho_{out} = R_z(\theta)\rho_{in}R_z(\theta)^\dagger$. In other words, the gate teleportation circuit implements the $R_z(\theta)$ operation on the computational qubit $\rho_{in}$.

Based on the circuit implementation in Fig.~\ref{figs2}, the specific form of the gate teleportation circuit $\mathcal{G}$ implementing $R_z(\theta)$ can be written as  
\begin{equation}\label{s27}
\begin{aligned}
\mathcal{G}(\rho)=\Pi_+ U\rho U^\dagger \Pi_+ +
R_z(2\theta)\Pi_- U\rho U^\dagger \Pi_-R_z(2\theta),
\end{aligned}
\end{equation} 
where $U$ is the CNOT gate between two qubits, $\Pi_\pm = (I\pm Z_2)/2$ are the measurement operators on the second qubit. When the output rotation state is replaced by $X\ket{\theta}=\ket{-\theta}$, the gate teleportation circuit must be correspondingly modified to  
\begin{equation}\label{s28}
\begin{aligned}
\mathcal{G'}(\rho)=R_z(2\theta)\Pi_+ U\rho U^\dagger \Pi_+R_z(2\theta)+\Pi_- U\rho U^\dagger \Pi_-.
\end{aligned}
\end{equation}
In practice, transforming $\mathcal{G}$ into $\mathcal{G}'$ only requires modifying the measurement feedback condition for executing the $R_z(2\theta)$ gate on the first qubit.

Next, we need to prove that for a noisy rotation state  
\begin{equation}
\begin{aligned}
\tilde{\rho}= (1-\epsilon) \ket{\theta}\bra{\theta} + \epsilon \ket{\theta^\perp}\bra{\theta^\perp} + b\ket{\theta}\bra{\theta^\perp} + b^*\ket{\theta^\perp}\bra{\theta}
\end{aligned}
\end{equation}
executing $\mathcal{G}$ and $\mathcal{G}'$ with equal probability (1/2 each) eliminates the effect of off-diagonal terms while preserving the effect on diagonal terms. It is straightforward to verify that $\mathcal{G}$ and $\mathcal{G}'$ as Eq.~(\ref{s27}) and Eq.~(\ref{s28}) satisfy the requirements of the gate teleportation circuit, i.e.,  
\begin{equation}
\begin{aligned}
&\mathcal{G}(\rho_{in} \otimes \ket{\theta}\bra{\theta})=\rho_{out} \otimes I,\\
&\mathcal{G'}(\rho_{in} \otimes (X_2\ket{\theta}\bra{\theta}X_2))=\rho_{out} \otimes I,
\end{aligned}
\end{equation}
where $\rho_{out} = R_z(\theta)\rho_{in}R_z(\theta)^\dagger$.

For the other diagonal term $\ket{\theta^\perp}\bra{\theta^\perp}$, after applying $\mathcal{G}$ and $\mathcal{G}'$, we obtain  
\begin{equation}
\begin{aligned}
&\mathcal{G}(\rho_{in} \otimes \ket{\theta^\perp}\bra{\theta^\perp}) 
= \mathcal{G}(\rho_{in} \otimes (Z_2\ket{\theta}\bra{\theta}Z_2))
= Z_1\otimes Z_2 \mathcal{G}(\rho_{in} \otimes \ket{\theta}\bra{\theta})Z_1\otimes Z_2
= (Z_1\rho_{out}Z_1) \otimes I.
\end{aligned}
\end{equation}
Here, we utilize the commutation relation  
\begin{equation}
\begin{aligned}
\mathcal{G}(Z_2\rho Z_2)=Z_1\otimes Z_2\mathcal{G}(\rho)Z_1\otimes Z_2,
\end{aligned}
\end{equation}
which can be verified via Eq.~(\ref{s27}). Similarly, we obtain  
\begin{equation}
\begin{aligned}
\mathcal{G'}(\rho_{in} \otimes X\ket{\theta^\perp}\bra{\theta^\perp}X) 
= (Z_1\rho_{out}Z_1) \otimes I.
\end{aligned}
\end{equation}
This demonstrates that neither $\mathcal{G}$ nor $\mathcal{G}'$ alters the effect on diagonal terms.

On the other hand, consider the off-diagonal term $\ket{\theta^\perp}\bra{\theta}= Z\ket{\theta}\bra{\theta}$. Similarly, the commutation relation is  
\begin{equation}
\begin{aligned}
&\mathcal{G}(Z_2\rho)=Z_1\otimes Z_2\mathcal{G}(\rho)\\
&\mathcal{G}'(Z_2\rho)=Z_1\otimes Z_2\mathcal{G}(\rho)
\end{aligned}
\end{equation}  
Thus, the corresponding output for this term is  
\begin{equation}
\begin{aligned}
&\mathcal{G}(\rho_{in} \otimes \ket{\theta^\perp}\bra{\theta}) 
= \mathcal{G}(\rho_{in} \otimes (Z_2 \ket{\theta}\bra{\theta})) 
= (Z_1\rho_{out}) \otimes I,\\
&\mathcal{G'}(\rho_{in} \otimes (X_2\ket{\theta^\perp}\bra{\theta}X_2)) 
=\mathcal{G'}(\rho_{in} \otimes (X_2Z_2\ket{\theta}\bra{\theta}X_2)) 
= -(Z_1\rho_{out}) \otimes I,
\end{aligned}
\end{equation}
where the negative sign in the second term arises from the anti-commutation of $Z_2$ and $X_2$. Therefore, executing $\mathcal{G}$ and $\mathcal{G}'$ with equal probability cancels the effect of the off-diagonal term. In the above discussion, we assumed that the subsequent correction operation $R_z(2\theta)$ can also be perfectly implemented. Executing $R_z(2\theta)$ consumes the rotation state $\ket{2\theta}$. Applying a similar randomization process to state $\ket{2\theta}$ iteratively eliminates the effect of off-diagonal terms in subsequent correction operations required for implementing the rotation gate. 

Thus, we have proven the effectiveness of the method for eliminating off-diagonal terms as described in the main text. For a single-qubit quantum state $\rho$ without off-diagonal terms, it can be verified that the infidelity between $\rho$ and the target state $\ket{\theta}$ equals the trace distance. Here, the infidelity and trace distance are defined respectively as:
\begin{equation}
\begin{aligned}
&1 - F(\rho, \ket{\theta}) = 1 - \langle \theta | \rho | \theta \rangle,\\
&D(\rho, |\theta\rangle\langle\theta|) = \frac{1}{2} \left\| \rho - |\theta\rangle\langle\theta| \right\|_1.
\end{aligned}
\end{equation}

\section{Surface-code implementation of multi-level transversal injection and state distillation}
\subsection{Implementation of Multi-Level Transversal Injection}

In this section, we provide a detailed description of how to implement the MLTI protocol on surface codes, and analyze the space-time volume. The first level of the MLTI protocol is the physical-level transversal injection. We consider implementing the multi-rotation transversal injection protocol from Ref.~\cite{PhysRevX.15.021057}. Compared to the original transversal injection protocol~\cite{choi2023faulttolerantnoncliffordstate}, the multi-rotation transversal injection protocol offers a significant advantage in post-selection success rate (and consequently in overhead). We employ three-qubit-rotations, i.e., \( m = 3 \), as larger \( m \) yields higher post-selection success rates. To execute three-qubit rotation gates on data qubits, we use the circuit Ref.~\cite{PhysRevX.15.021057}, utilizing syndrome qubits between data qubits as ancillas.

Here, logical qubits are encoded in a rectangular array of rotated surface codes, where the logical \( X \) and logical \( Z \) code distances satisfy \( d_{z}^{(1)} > d_{x}^{(1)} \). To suppress measurement errors on syndrome qubits to \( \mathcal{O}(p_{phy}^2) \), two rounds of syndrome extraction circuits are repeated. Subsequently, only cases where the measurement results of the first two rows of \( X \)-type syndrome qubits are both \( +1 \) in both rounds are retained.

To fix the phase of the output state to \( +i \), an \( S \) or \( S^\dagger \) gate needs to be applied when \( k^{(1)} = d_{z}^{(1)}/m \) is even. However, for the MLTI with level two or more, this step can be merged into the implementation of the \( R_z(-\pi/8) \) gate during magic state pumping, thus incurring no additional overhead. That is to say, we only account for the overhead of this step in separate physical-level protocols. Thus, the space-time volume of the first-level MI protocol can be calculated as  
\begin{equation}
\begin{aligned}
V^{(1)}=2d_{z}^{(1)}d_{x}^{(1)} \times 2 /p_{s,1},
\end{aligned}
\end{equation} 
where \( p_{s,1} \) is the post-selection success rate of the first-level protocol and the space overhead of the rectangular rotated surface code array is estimated as \( 2d_{z}^{(1)}d_{x}^{(1)} \), including data qubits and syndrome qubits.

\begin{figure*}[tpb]
\centering
\includegraphics[width=14cm]{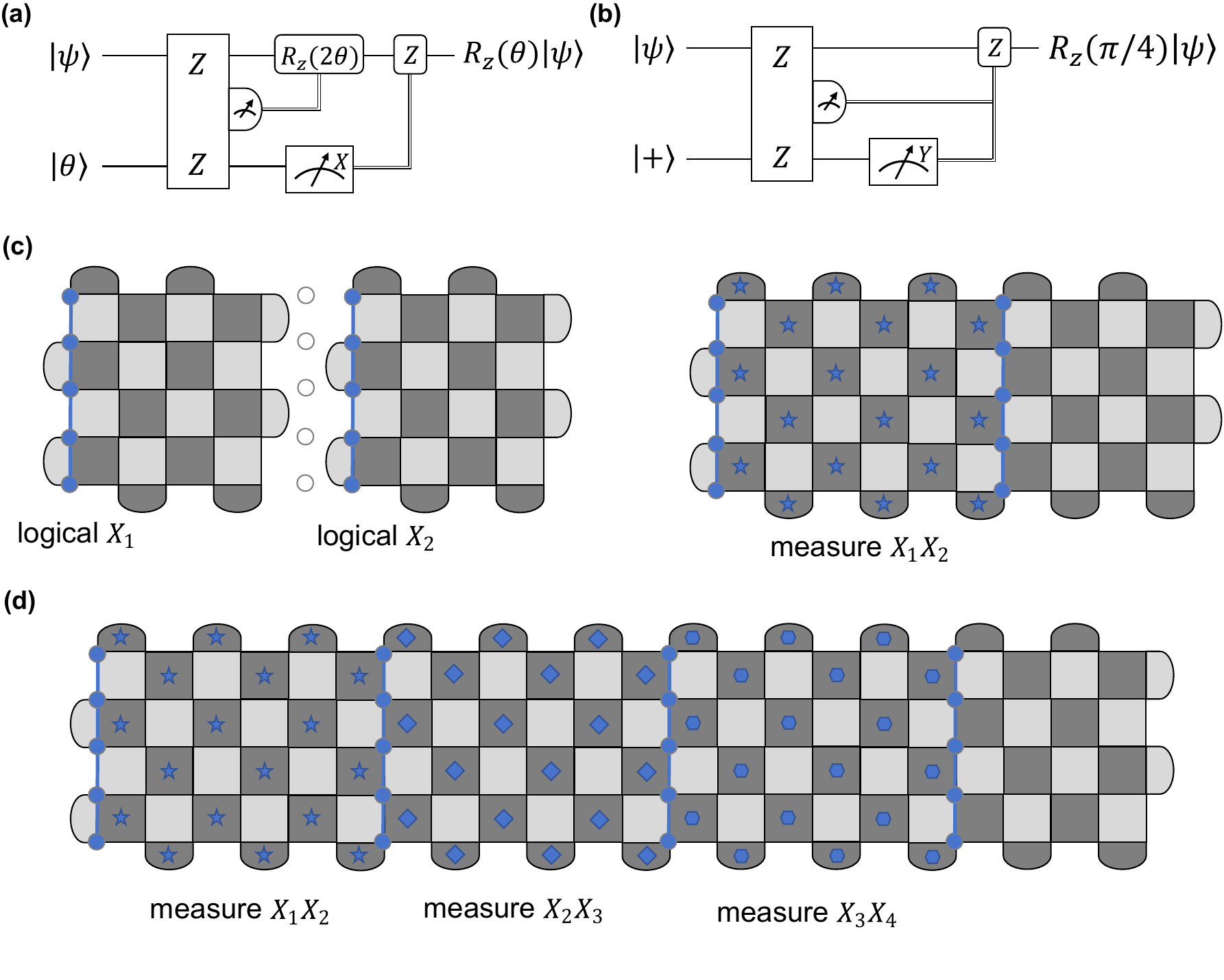}
\caption{\textbf{Lattice surgery operations involved in the MLTI protocol.  }
\textbf{(a-b)} Implementation of \(R_z(\theta)\) and \(R_z(\pi/4)\) gates by Pauli-basis measurements, which are supported by lattice surgery on the surface code.  
\textbf{(c)} Measurement of the operator \(X_1 X_2\) on two logical qubits by surface code lattice surgery. The outcome of the \(X_1 X_2\) measurement is read out by the product of all star stabilizer measurement values (marked with asterisks). 
\textbf{(d)} Parallel measurement of the operator \(X_i X_{i+1}\) on multiple surface code logical qubits. Since the readout regions do not overlap, the lattice surgery procedures can be performed in parallel. After patch merging, defining the new logical \(X_L = X_1\) results in a concatenated code composed of a surface code and a one-dimensional repetition code. This concatenated code is itself a larger surface code.
}\label{figs3}
\end{figure*}

Next, to perform magic state pumping, a magic state with fidelity comparable to the output state of is consumed. Assume the magic state \( \ket{\pi/8} \) is prepared in a square surface code of size \( d_{z}^{(1)} \times d_{z}^{(1)} \). This logical qubit can interact with the first-level output state of size \( d_{x}^{(1)} \times d_{z}^{(1)} \) with lattice surgery by measuring \( Z \otimes Z \), thereby implementing the \( R_z(- \pi/8) \) gate (see Fig.~\ref{figs3}a). To ensure fault tolerance, assume the lattice surgery process spans \( d_{z}^{(1)} \) QEC cycles. During the execution of \( R_z(- \pi/8) \), there is a 1/2 probability that an \( R_z(\pm \pi/4) \) gate (i.e., \( S \) or \( S^\dagger \) gate) needs to be applied. This can be achieved by preparing a \( \ket{0} \) state of size \( d_{z}^{(1)} \times d_{z}^{(1)} \) and then executing the circuit in Fig.~\ref{figs3}b. The \( Y \)-basis measurement in this circuit can be implemented using the method from Ref.~\cite{Gidney_2024}, lasting \( \lfloor d_{z}^{(1)}/2 \rfloor + 2\) cycles. Overall, the space-time volume for the magic state pumping stage can be calculated as  
\begin{equation}
\begin{aligned}
V^{(1)}_{pump}= V^{(1)}_{T} + 2 d_{z}^{(1)}(d_{x}^{(1)}+d_{z}^{(1)}+1)\times (2 d_{z}^{(1)} + \lfloor d_{z}^{(1)}/2 \rfloor+2),
\end{aligned}
\end{equation}
where \( V^{(1)}_{T} \) is the space-time volume for preparing a magic state using cultivation~\cite{gidney2024magicstatecultivationgrowing} or distillation~\cite{Gidney2019efficientmagicstate} protocols, \( 2 d_{z}^{(1)}(d_{x}^{(1)}+d_{z}^{(1)}+1) \) is the number of physical qubits occupied during lattice surgery, and \( (2 d_{z}^{(1)} + \lfloor d_{z}^{(1)}/2 \rfloor) +2)\) is the total number of cycles for executing \( R_z(- \pi/8) \) and \( R_z(\pm \pi/4) \) gates.

The second-level protocol starts from \( k_2 \) output states from the first-level protocol. Through patch deformation, the size of each logical qubit can be modified to \( d_x^{(2)} \times d_z^{(2)} \). To perform syndrome measurements on the higher-level concatenated code, we utilize lattice surgery to measure the \( X_iX_{i+1} \) operator between adjacent surface code logical qubits. This process can be implemented within the merge operation of lattice surgery, see Fig.~\ref{figs3} (c-d). Here, to ensure fault tolerance of the measurement, we set the number of QEC rounds to \( d_z^{(2)} + 2 \) to suppress logical measurement errors.

Subsequently, only cases with trivial measurement results are retained, and the merged larger code block is defined as a new logical qubit of surface codes with size \( d_x^{(2)} \times (k^{(2)}d_z^{(2)} + k^{(2)} - 1) \). Overall, the space-time volume for the two-level protocol is  
\begin{equation}
\begin{aligned}
V^{(2)} = [2 d_x^{(2)} (k^{(2)} d_z^{(2)} + k^{(2)} -1) \times (d_z^{(2)} + 2) +V^{(1)} + V^{(1)}_{pump}]
/p_{s,2}
\end{aligned}
\end{equation}
where \( p_{s,2} \) is the post-selection success rate of the second-level protocol. By repeating the above discussion, the space-time volume of the MLTI protocol on any level can be derived.

Next, we analyze the logical error rates of each component. A basic assumption is that the per-cycle memory error rate of a logical $X$ or $Z$ on a surface code patch of size $d_x \times d_z$ is denoted as $P^{x/z}_L(d_x, d_z)$. First, during the two-cycle syndrome measurement at the first level, the logical error rate is estimated as 
\begin{equation}
\begin{aligned}
2 P^{x/z}_L(d_x^{(1)}, d_z^{(1)}).
\end{aligned}
\end{equation}
Second, in the magic state pumping stage (level 1 to level 2), the logical error rate of performing the $\pi/8$ rotation gate arises from three sources. This includes the inherent error of the magic state, and the memory error during the lattice merger phase, estimated as 
\begin{equation}
\begin{aligned}
d_z^{(1)} \times P^{x/z}_L(d_x^{(1)}+d_z^{(1)}+1, d_z^{(1)}),
\end{aligned}
\end{equation}
as well as measurement error. For the measurement error, a conservative estimate is that its probability is no greater than the memory error during the lattice surgery process, which is supported by results from Ref.~\cite{Gidney_2022} and \cite{PRXQuantum.3.010331}. Likewise, if a $\pi/4$ rotation gate needs to be executed, the memory error during the lattice surgery phase and the measurement error must be considered. Furthermore, since the $Y$ measurement lasts for $d_z^{(1)}/2$ cycles, the logical qubit idle error is 
\begin{equation}
\begin{aligned}
d_z^{(1)}/2 \times P^{x/z}_L(d_x^{(1)}, d_z^{(1)}). 
\end{aligned}
\end{equation}
If the qubit does not need to perform a $\pi/4$ rotation gate, the logical qubit must remain idle for a total of $2.5d_z^{(1)}/2$ cycles, with an error rate of 
\begin{equation}
\begin{aligned}
2.5\times d_z^{(1)}/2 \times P^{x/z}_L(d_x^{(1)}, d_z^{(1)}).
\end{aligned}
\end{equation}

Similarly, during transversal injection on level $r$, the error originates from lattice surgery operations performed on larger code patches. The memory error rate is 
\begin{equation}
\begin{aligned}
(d_z^{(r)}+2) \times P^{x/z}_L(d_x^{(r)} + d_z^{(r)} +1, d_z^{(r)}).
\end{aligned}
\end{equation}
As for the measurement error rate, the number of measurement rounds is set to $d_z^{(2)} + 2$. Under our physical error rate settings ($p_{phy}\leq 10^{-3}$), the measurement error will be over two orders of magnitude lower than the memory error. Therefore, we neglect its contribution.

\subsection{Implementation of State Distillation}
\begin{figure*}[htpb]
\centering
\includegraphics[width=17.5cm]{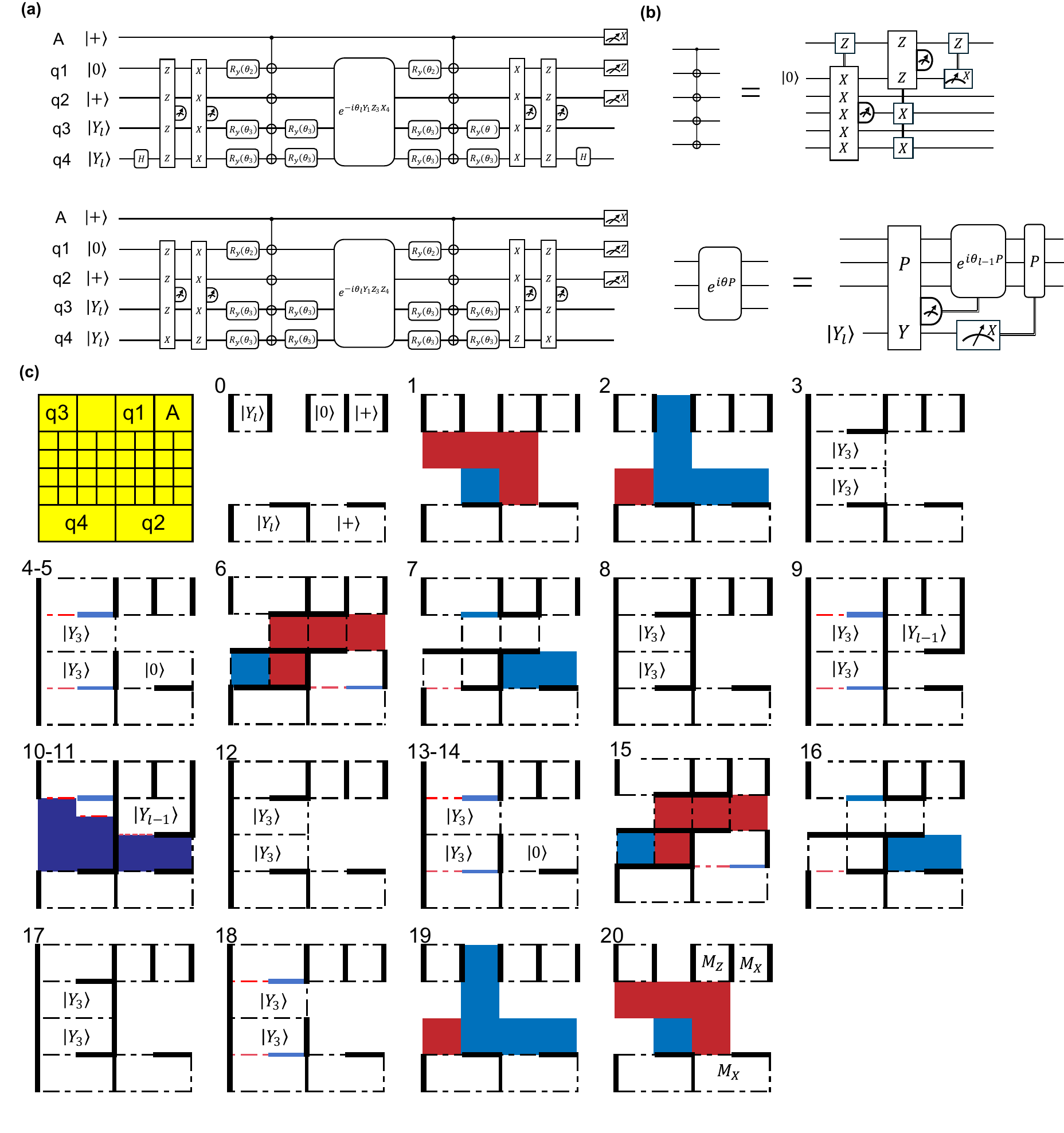}
\caption{ \textbf{Surface code implementation of the state distillation protocol.  }
\textbf{(a)} Compilation of the state distillation circuit from Fig. S1c based on Pauli-basis measurements. The encoding and decoding processes are replaced by Pauli-basis measurements of \(XXXX\) and \(ZZZZ\). The intermediate single-qubit rotation gates and surrounding Clifford operations are combined into multi-qubit rotation gates. The bottom circuit is derived by moving the H gates in the top circuit to the end.  
\textbf{(b)} Implementation of multi-target CNOT gates and rotation gates in the circuit.  
\textbf{(c)} Realization of the distillation circuit under a surface code layout. Dashed and bold solid lines represent the logical Z-boundary and logical X-boundary of the surface code block, respectively. The side length of the minimal unit cell is \(d/2\). The protocol occupies approximately \(32d^2\) physical qubits in area and lasts for \(20d\) cycles. The whole process is described as follows. 0. Initialize code blocks.  
1–2. Measure $ZZZX$ and $XXXZ$ to encode the [[4,2,2]] code. 
3. Patch deformation and transport of $\ket{Y_3}$.  
4–5. Perform $R_y(\theta_3)$ and $R_y(\theta_2)$ gates by lattice surgery.  
6–7. Perform $ZZZZ$ and $XX$ measurements to implement the multi-target CNOT gate.  
8. Transport $\ket{Y_3}$.  
9. Perform $R_y^\dagger(\theta_3)$ rotation by lattice surgery while transporting $\ket{Y_{l-1}}$.  
10–11. Perform the $(l-1)$-level multi-qubit rotation $e^{-i\theta_{l-1}Y_1Z_3Z_4}$.  
12–17. Same as steps 3–8.  
18. Perform $R_y^\dagger(\theta_2)$ rotations by lattice surgery.  
19–20. Measure $XXXZ$ and $ZZZX$, then measure the auxiliary state and qubits 1 and 2.}\label{figs4}
\end{figure*}

For comparison, we also consider the implementation of the state distillation circuit~\cite{Campbell_2016} in the surface code framework. Specifically, lattice surgery on the surface code are used to execute the circuit shown in Fig.~\ref{figs1}c. To better adapt the original circuit to the native operations of lattice surgery, we made several modifications to the distillation circuit, as illustrated in Fig.~\ref{figs4}a.

First, we replaced the encoding and decoding processes at the beginning and end of the circuit with Pauli-basis measurements (up to Pauli operations), which were constructed from a series of Clifford gates in the original reference~\cite{PhysRevA.91.042315,Campbell_2016}. Second, we merged the $(l-1)$-level single-qubit rotation gate in the middle of the circuit and their surrounding Clifford operations into multi-qubit rotation gate $e^{-i\theta_{l-1}Y1Z3X4}$, which are also supported by lattice surgery. Third, we moved the $H$ gate on qubit 4 to the end of the circuit. Finally, the correction operation $R_y(\theta_2)$ after the $R_y(\theta_3)$ gate in layer 2 was merged with subsequent Clifford operations.

Fig.~\ref{figs4}b explicitly shows the qubit layout we designed and the operations performed in each time step. Overall, one execution of the state distillation protocol takes $20d$ cycles and requires a total of $2 \times 16d^2$ physical qubits, where $d$ is the code distance of the qubit 1 (or qubit 3) in the Fig.~\ref{figs4}b.

Next, let us analyze in detail the impact of errors on each qubit on the distillation outcome. For the ancilla qubit, both $Z$ and $X$ errors can be detected, resulting in a protocol failure rate of \(20 \times (P_L^z(d,d) + P_L^x(d,d))\). For qubit 1, all $X$ errors are detectable, leading to a failure rate of \(20 \times P_L^x(d,d)\), while $Z$ errors on qubit 1 have no effect outside the 4-qubit code and are fully detectable within the code, contributing a failure rate of \(18 \times P_L^z(d,d)\). Qubit 2 exhibits the same behavior as qubit 1. Note that qubit 2 is encoded on an asymmetric rectangular surface code, making its error rate less straightforward to estimate. Nevertheless, as an optimistic estimate, the error rate on qubit 2 will be no lower than that on qubit 1. For qubit 3, a $Z$ error in the final step contributes \(P_L^z(d,d)\) to the output infidelity; both $X$ and $Z$ errors within the encoded space are detectable, resulting in a failure rate of \(20 \times P_L^x(d,d) + 18 \times P_L^z(d,d)\). Qubit 4 exhibits the same behavior as qubit 3. Similarly, the error rate on qubit 4 will be no lower than that on qubit 3. Additionally, in this layout, the logical error rate in the routing region is negligible as its width is larger than the code distance $d$.

Overall, the optimistic estimate of failure probability of the distillation protocol is:  
\begin{equation}
\begin{aligned}
p_{fail} \simeq 92P_L^z(d,d)+100P_L^x(d,d) + 8\epsilon_3 + 2\epsilon_l +0.5\eta_{l-1},
\end{aligned}
\end{equation}
where $\epsilon_3$, $\epsilon_l$, and $\eta_{l-1}$ represent the error rates of $\ket{Y_3}$, $\ket{Y_l}$, and ${R_y(\theta_{l-1})}$, respectively~\cite{Campbell_2016}. The output error rate of the protocol is:  
\begin{equation}
\begin{aligned}
\epsilon_{out} \simeq P_L^z(d,d) + 8\epsilon_3^2 + \epsilon_l^2 +0.25\eta_{l-1},
\end{aligned}
\end{equation} 
Thus, the space-time overhead averaged over the two output states is
\begin{equation}
\begin{aligned}
16d^2 \times 20d / (1-p_{fail} ).
\end{aligned}
\end{equation} 

\section{Details of numerical simulations}

\subsection{Simulation Method}

Here, we describe the simulation approach for transversal injection. Assume the matrix representation of the input state $\rho_{\text{in}}$ on each qubit in the $\ket{+}, \ket{-}$ basis is given by:
\begin{equation}
\begin{aligned}
\rho_{in}= 
\begin{pmatrix}
\rho_{++} & \rho_{+-}\\ \rho_{-+} & \rho_{--}
\end{pmatrix},
\end{aligned}
\end{equation} 
where the parameters $\rho_{ij}$ satisfy the requirements for a physical density matrix. After an ideal transversal injection, only the following component from the input state ${\rho_{in}}^{\otimes k}$ can be post-selected:
\begin{equation}
\begin{aligned}
\rho_{++}^k\ket{+^{\otimes k}}\bra{+^{\otimes k}},\quad
\rho_{+-}^k\ket{+^{\otimes k}}\bra{-^{\otimes k}},\\
\rho_{-+}^k\ket{-^{\otimes k}}\bra{+^{\otimes k}},\quad
\rho_{--}^k\ket{-^{\otimes k}}\bra{-^{\otimes k}}.
\end{aligned}
\end{equation} 
Thus, under the definition of new bases $\ket{\pm_L} = \ket{\pm^{\otimes k}}$, the density matrix of the output state becomes:
\begin{equation}
\begin{aligned}
\rho_{out}= \frac{1}{\rho_{++}^k+\rho_{--}^k}
\begin{pmatrix}
\rho_{++}^k & \rho_{+-}^k\\ \rho_{-+}^k & \rho_{--}^k
\end{pmatrix}.
\end{aligned}
\end{equation} 
From this, the fidelity between the output state and the target state can be computed.

To evaluate the effect of noise, we introduce noise channels on the input state $\rho_{\text{in}}$. Specifically, for logical qubits, we consider the following Pauli noise channel:
\begin{equation}
\begin{aligned}
\mathcal{E}_{L}(\rho)&=(1-P_L^x-P_L^z)\rho_1 + P_L^x X \rho X + P_L^z Z \rho Z.
\end{aligned}
\end{equation}
In fact, all errors on the logical qubits, including memory and measurement errors, can be equivalently represented as Pauli $X$ or $Z$ errors at certain locations in the circuit. 

For surface codes with an odd code distance under the circuit-level noise model, the logical error rate can be estimated using the following formula~\cite{PhysRevA.86.032324}
\begin{equation}
\begin{aligned}
P_L (d) =0.1(100p_{phy})^{(d+1)/2},
\end{aligned}
\end{equation}
where logical \(X_L\) and \(Z_L\) errors each account for half, i.e., 
\begin{equation}
\begin{aligned}
P_L^x(d,d)=P_L^z(d,d)=0.05(100p_{phy})^{(d+1)/2},
\end{aligned}
\end{equation}
To estimate the logical error rate on rectangular surface codes, the fundamental relationship used is that the number of shortest error chains causing a logical error is approximately proportional to the area of the surface code region. Therefore, the logical error rate for rectangular surface codes with odd code distances can be estimated as~\cite{Litinski2019magicstate}
\begin{equation}
\begin{aligned}
&P_L^x(d_x,d_z)=\frac{d_xd_z}{d_x^2}P_L^x(d_x,d_x),\\
&P_L^z(d_x,d_z)=\frac{d_xd_z}{d_z^2}P_L^x(d_z,d_z).
\end{aligned}
\end{equation}
For surface codes with an even code distance \(d\), the length of the shortest error chain causing a logical error is \(d/2\), which is consistent with that of a surface code with code distance \(d-1\). Due to symmetry, it is reasonable to assume that half of these error chains can be corrected by the surface code with even code distance \(d\). Hence, the logical error rate for rectangular surface codes with even code distances can be estimated as
\begin{equation}
\begin{aligned}
&P_L^x(d_x,d_z)=\frac{d_xd_z}{2(d_x-1)^2}P_L^x(d_x-1,d_x-1),\\
&P_L^z(d_x,d_z)=\frac{d_xd_z}{2(d_z-1)^2}P_L^x(d_z-1,d_z-1).
\end{aligned}
\end{equation}

For physical qubits at the first level, we also simulate the circuit performing multi-qubit rotations and two surface code error-correction cycles. In the simulation, we adopt a standard circuit-level noise model. Specifically, the noise channels are defined as follows:
\begin{equation}
\begin{aligned}
\mathcal{E}_{1}(\rho_1)&=(1-p)\rho_1+(p/3)\sum_{P\in\{X,Y,Z\}}P\rho_1P,\\
\mathcal{E}_{2}(\rho_2)&=(1-p)\rho_2+(p/15)
\times \sum_{\substack{P_1,P_2\in\{I,X,Y,Z\},\\P_1\otimes P_2\neq I \otimes I}}P_1\otimes P_2\rho_2P_1\otimes P_2,
\end{aligned}
\end{equation}
where $\rho_1$ and $\rho_2$ are single-qubit and two-qubit density matrices, respectively. In the simulated circuits, we apply $\mathcal{E}_1$ after single-qubit gates and idle operations, and $\mathcal{E}_2$ after two-qubit gates. Additionally, measurement outcomes and state initializations flip with probability $p$.

For the multi-rotation transversal injection protocol with $m = 3$, if single- or two-qubit $Z$ errors occur on the $k$ physical qubits, the result cannot pass the post-selection. Based on simulation results, we obtain the probability of discarding these outcomes after two cycles. Furthermore, results from Ref.~\cite{PhysRevX.15.021057} indicate that the probability of undetectable 3-qubit $Z$ errors on each set of physical qubits is $\frac{2}{15}p_{{phy}} + \mathcal{O}(p_{phy}^2)$. Using these results, we introduce an equivalent Pauli noise channel for the initial physical qubits.

\subsection{Search Method}  
To minimize the space-time volume of the MLTI protocol under a target fidelity, it is necessary to search over parameters at each level of the protocol. Specifically, these parameters include the code distances $d_x^{(r)}$ and $d_z^{(r)}$ on each level (denoted by $r$), the number of input states $k^{(r)}$, and the parameters of the magic states consumed on each level. For high-level MLTI protocols, an exhaustive search over the entire parameter space is infeasible. Therefore, we employ the following strategies to narrow down the search space.

First, the number of input states $k^{(r)}$ at each level directly determines the angular relationship between the output and input states. For a target output angle, we prefer the input angle to be near $\pi/8$ to facilitate preparation by magic state pumping. Thus, for an output angle $\beta$, the values of $k^{(r)}$ within the search range should satisfy:
\begin{equation}
\begin{aligned}
\alpha = \arctan(({\tan \beta})^{1/k^{(r)}}) \in (-\delta, \delta),
\end{aligned}
\end{equation}
where we set $\delta = \pi/16$.

Second, the code distances $d_x^{(r)}$ and $d_z^{(r)}$ at each level can be searched independently. Note that the objective function (space-time volume) is strictly monotonic with respect to the code distances. By fixing other code distances to sufficiently large values, we can perform an independent local search for each code distance and narrow the range near its optimal value.

Finally, we search for magic states with suitable fidelity to further reduce the space-time volume. The magic states used in this work are sourced from two protocols. For cases where high precision is not critical, we use the cultivation protocol from Ref.~\cite{gidney2024magicstatecultivationgrowing}. For higher precision requirements, we apply an additional round of distillation protocol~\cite{Gidney2019efficientmagicstate} to the output states of the cultivation protocol. These results are applied equitably to MLTI, state distillation, and gate synthesis protocols. 

\begin{figure}[htbp]
\centering
\includegraphics[width=10cm]{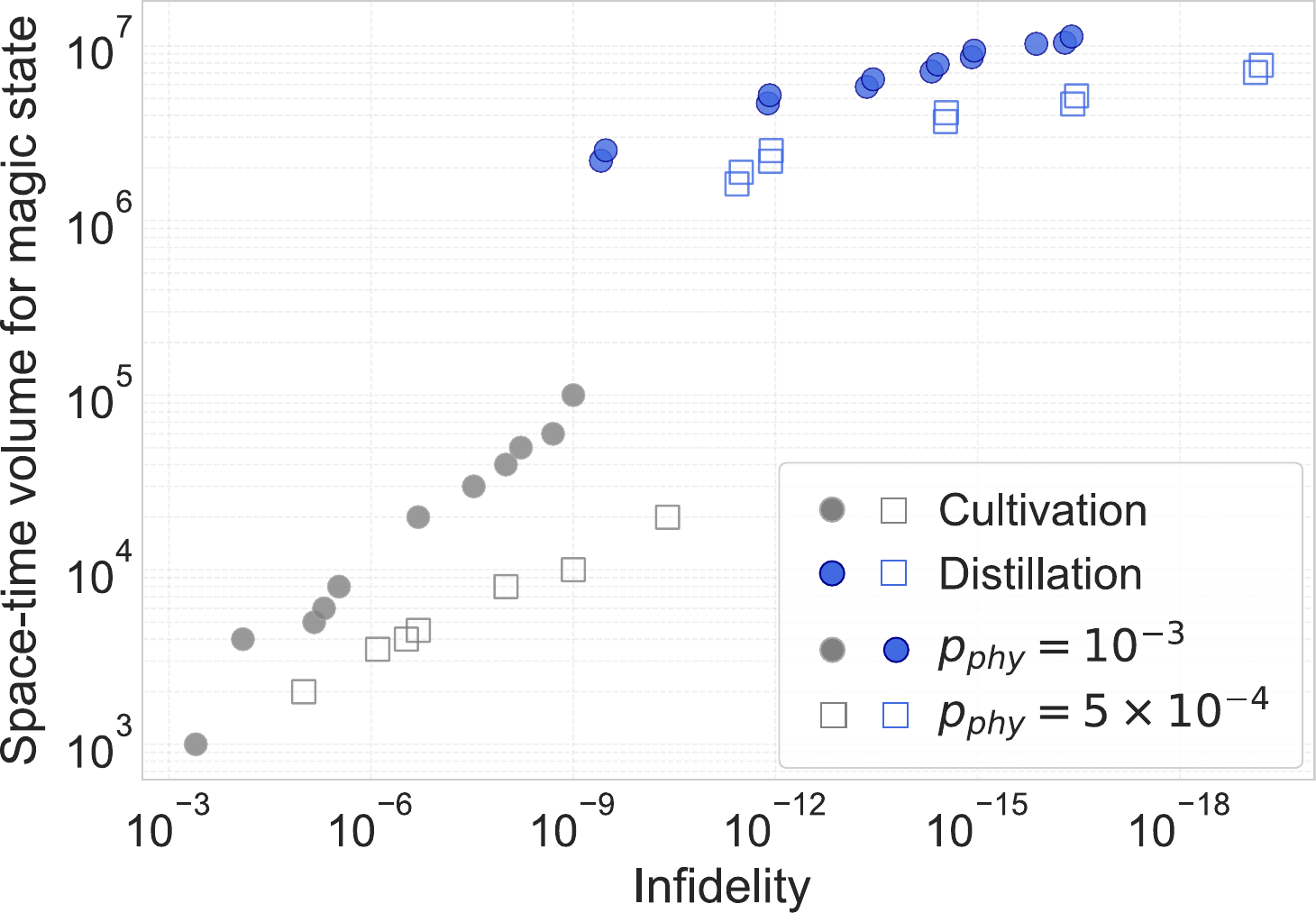}
\caption{ \textbf{Space-time volumes for preparing magic state used in this work.} The cultivation protocol is used to prepare magic states with relatively moderate fidelity requirements, where data are from Ref.~\cite{gidney2024magicstatecultivationgrowing}. The distillation protocol takes the output states of cultivation as input and is used to prepare magic states with higher fidelity.}\label{figs5}
\end{figure}

We evaluated the output fidelity and space-time volumes of these protocols for preparing magic states (see Fig.~\ref{figs5}) and used these data in the search. To constrain the search, we use the expression
\begin{equation}
\begin{aligned}
\epsilon_{L,T} \simeq k^{(r)} \epsilon_T \beta^{2(1 - 1/k^{(r)})}
\end{aligned}
\end{equation}
to estimate the impact of the magic state infidelity $\epsilon_T$ on the output state fidelity, where $\epsilon_T$ is the infidelity of the magic state and $\beta$ denotes the output angle at the current level. We focus the search near values of $\epsilon_{T}$ for which $\epsilon_{L,T}$ is near the target infidelity.

\subsection{Other Details}
For Fig.~2 in the main text, all code distance parameters \(d_x^{(r)}\) and \(d_z^{(r)}\) are set sufficiently large such that all logical error rates approach zero. Similarly, the error rates of all magic states are set to zero. Under these settings, the parameters \(k^{(r)}\) are optimized to achieve the optimal fidelity.

For Fig.~3a, all parameters are optimized to minimize the space-time overhead under the constraint that the output infidelity is below the target infidelity. For the compared state distillation protocols, the code distances of the protocols and the parameters of the magic states are optimized to minimize the space-time overhead. For a target infidelity of \(10^{-12}\), a single-level distillation protocol is insufficient. Therefore, a two-level distillation is employed, with identical code distances and magic state parameters used on each level.

For Fig.~3b, the overhead of implementing rotation gates using the rotation states from Fig.~3a is considered. Here, the rotation gates are executed sequentially. That is, \(R_z(\theta)\) is executed first, with a 50\% probability of proceeding to execute \(R_z(2\theta)\), followed by a 25\% probability of executing \(R_z(4\theta)\), and so on. The space-time volume shown in the figure includes only the volume cost of preparing these rotation states, i.e.,  
\begin{equation}
\begin{aligned}
V_{gate,l} =\sum_{i=3}^l V_l / 2^{l-i},
\end{aligned}
\end{equation} 
where \(V_l\) denotes the space-time volume used for either the MLTI protocol or the state distillation protocol to prepare a rotation state at Clifford hierarchy level \(l\), whichever one is smaller. For comparison, the volume estimate for gate synthesis only accounts for the magic state preparation cost associated with the number of \(T\) gates, i.e.,  
\begin{equation}
\begin{aligned}
V'_{gate,l} = n_T \times V_T,
\end{aligned}
\end{equation}
where \(V_T\) is the space-time volume required to prepare one magic state and \(n_T\) is the magic state count. With this magic state count, the precision $\delta$ contributes to the error rate as ${\delta}^2$ since $1-|\braket{\alpha|\alpha+\delta}|^2=\sin^2\delta \simeq\delta^2$, where $\delta$ satisfies $n_T\simeq 3\log_2(1/\delta)$. To ensure that the ancilla states in both protocols introduce comparable error rates during gate implementation, the following relation must be satisfied:  
\begin{equation}
\begin{aligned}
n_T \times \epsilon_T + {\delta}^2  \leq 2 \times \epsilon_{target},
\end{aligned}
\end{equation} 
where \(\epsilon_{target}\) is the target infidelity of the rotation state produced by the MLTI protocol or state distillation, and \(\epsilon_T\) is the infidelity of the magic state in the gate synthesis protocol. Here, the factor of 2 comes from the fact that, on average, only one additional correction operation is needed to implement a rotation gate.

\section{Additional Numerical Results }
\begin{figure*}[htbp]
\centering
\includegraphics[width=18cm]{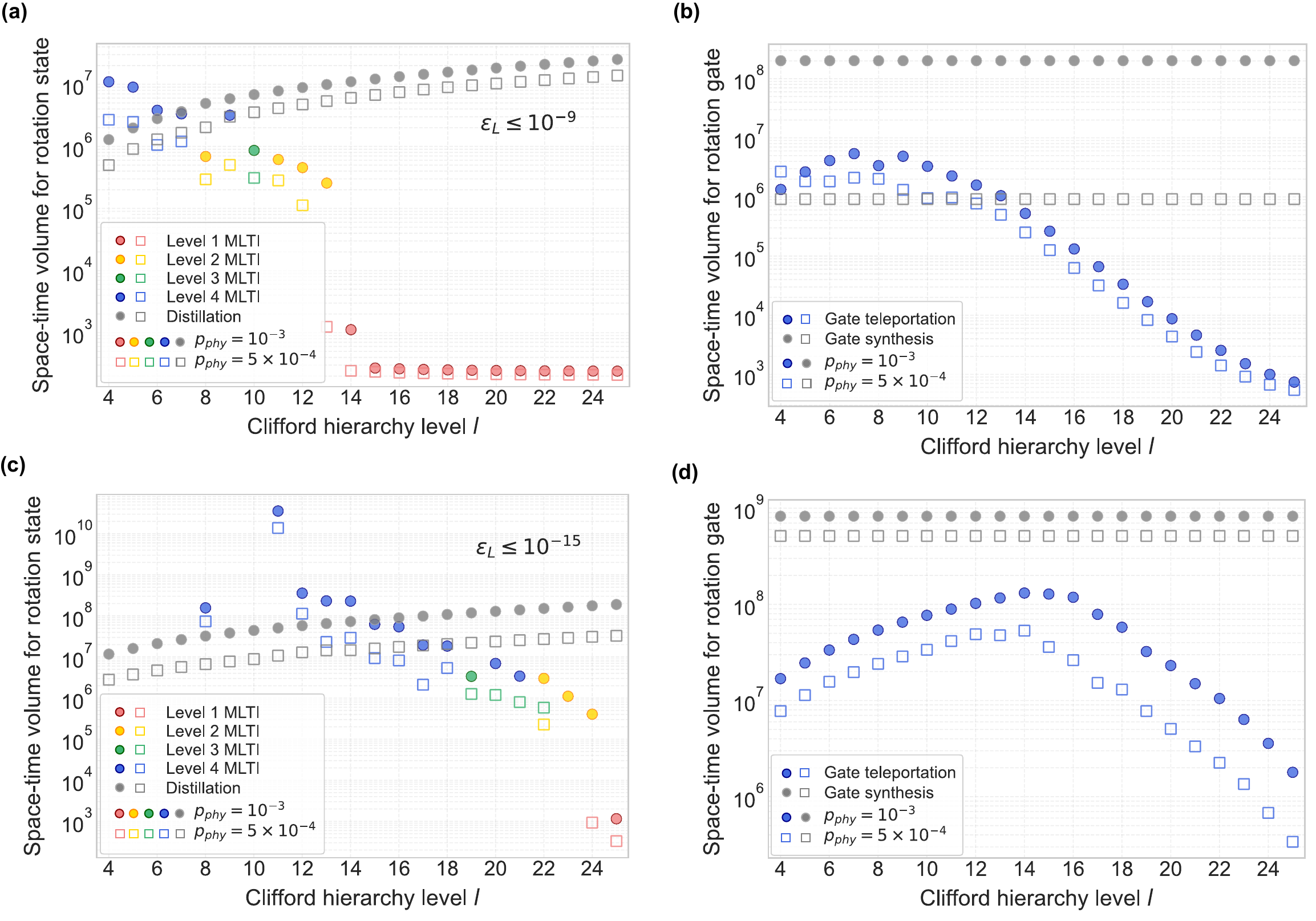}
\caption{\textbf{Additional numerical results.} \textbf{(a)} Space-time volume required to prepare rotation states using MLTI and state distillation, with a target infidelity of \(10^{-9}\).  
\textbf{(b)} Space-time volume required to implement rotation gates $R_z(\pi/2^l)$ by gate teleportation using the rotation states from (a), and the space-time volume required by gate synthesis to achieve rotation gates with comparable error rates. For the gate synthesis data, the significant performance difference between the $10^{-3}$ and $5\times 10^{-4}$ physical error rates arises because the $10^{-3}$ regime requires only the cultivation protocol, whereas the $5\times 10^{-4}$ regime necessitates costly distillation.
\textbf{(c)} Space-time volume required to prepare rotation states using MLTI and state distillation, with a target infidelity of \(10^{-15}\).  
\textbf{(d)} Space-time volume required to implement rotation gates $R_z(\pi/2^l)$ by gate teleportation using the rotation states from (c), and the space-time volume required by gate synthesis to achieve rotation gates with comparable error rates.}\label{figs6}
\end{figure*}

In the main text, we present the space-time volume required to prepare rotation states and perform rotation gates with a target state infidelity of $10^{-12}$. Here, we further provide results for target state infidelity of $10^{-9}$ and $10^{-15}$ in Fig.~\ref{figs6}. Similarly, we compare MLTI, state distillation, and gate synthesis. Roughly speaking, lower infidelity requires higher-level MLTI. Since we only implement MLTI up to level 4, the volume data of preparing rotation states for MLTI shows missing values or surges at $10^{-15}$ infidelity. We expect these cases will improve after parameter searches for higher-level MLTI. Moreover, under the gate synthesis scheme, for the case $p_{phy}=10^{-3}$, the optimal fidelity of the magic states used from Fig.~\ref{figs5} is insufficient to meet the fidelity requirements in Fig.~\ref{figs6}d. Therefore, for this case, Fig.~\ref{figs6}d shows a favorable estimate for gate synthesis at $p_{phy}=10^{-3}$, where $\epsilon_T=0$, taking the maximum space-time volume from Fig.~\ref{figs5}.

\end{document}